# Strain-Tuning of the Optical Properties of Semiconductor Nanomaterials by Integration onto Piezoelectric Actuators

*Javier Martín-Sánchez[1*], Rinaldo Trotta[1*], Antonio Mariscal[2], Rosalía Serna[2], Giovanni Piredda[3], Sandra Stroj[3], Johannes Edlinger[3], Christian Schimpf[1], Johannes Aberl[1], Thomas Lettner[1], Johannes Wildmann[1], Huiying Huang[1], Xueyong Yuan,[1,4] Dorian Ziss[1], Julian Stangl[1], and Armando Rastelli,[1,4*]*

[1] Institute of Semiconductor and Solid State Physics, Johannes Kepler University Linz, Altenbergerstrasse 69, A-4040 Linz, Austria

[2] Laser Processing Group, Instituto de Óptica, CSIC, C/Serrano 121, 28006 Madrid, Spain

[3] Research Center for Microtechnology, Vorarlberg University of Applied Sciences, 6850 Dornbirn, Austria

[4] Johannes Kepler University Linz, Linz Institute of Technology, Altenbergerstrasse 69, A-4040 Linz, Austria

The tailoring of the physical properties of semiconductor nanomaterials by strain has been gaining increasing attention over the last years for a wide range of applications such as electronics, optoelectronics and photonics. The ability to introduce deliberate strain fields with controlled magnitude and in a reversible manner is essential for fundamental studies of novel materials and may lead to the realization of advanced multi-functional devices. A prominent approach consists in the integration of active nanomaterials, in thin epitaxial films or embedded within carrier nanomembranes, onto $Pb(Mg_{1/3}Nb_{2/3})O_3$-$PbTiO_3$-based piezoelectric actuators, which convert electrical signals into mechanical deformation (strain). In this review, we mainly focus on recent advances in strain-tunable properties of self-assembled InAs quantum dots embedded in semiconductor nanomembranes and photonic structures. Additionally, recent works on other nanomaterials like rare-earth and metal-ion doped thin films, graphene and $MoS_2$ or $WSe_2$ semiconductor two-dimensional materials are also reviewed. For the sake of completeness, a comprehensive comparison between different procedures employed throughout the literature to fabricate such hybrid piezoelectric-semiconductor devices is presented. It is shown that unprocessed piezoelectric substrates (monolithic actuators) allow to obtain a certain degree of control over the nanomaterials' emission properties such as their emission energy, fine-structure-splitting in self-assembled InAs quantum dots and semiconductor 2D materials, upconversion phenomena in $BaTiO_3$ thin films or piezotronic effects in ZnS:Mn films and InAs quantum dots. Very recently, a novel class of micro-machined piezoelectric actuators have been demonstrated for a full control of in-plane stress fields in nanomembranes, which enables producing energy-tunable sources of polarization-entangled photons in arbitrary quantum dots. Future research directions and prospects are discussed.

6.- Conclusions

**1.- Introduction.**

Micro-electro-mechanical systems (MEMS) opened new avenues for an extended number of applications including ultrasound medical imaging, robotics, printing, photonics and energy harvesting [1,2]. Electrostatically actuated devices, commonly used in MEMS applications, present several drawbacks such as relatively high power consumption and substantial hysteresis. As a consequence, there has been an intensive research activity during the last decades in the fabrication of advanced piezoelectric materials and development of novel piezoelectric MEMS devices [3–7]. Piezoelectric materials have emerged as promising alternatives in MEMS for sensing and actuating applications, as they provide wide dynamic ranges, low power consumption and low hysteresis values [8]. The non-centrosymmetric arrangement of atoms inside the unit cell of such piezoelectric materials lead to electromechanical transduction phenomena that can be exploited for a wide range of sensing applications. In the converse regime, they respond to electrical signals by mechanical deformations that can be controlled accurately by fine tuning the excitation driving voltage and frequency. Piezoelectric materials can be machined in many different ways, such as cantilever-type structures, which can generate relatively large forces (actuation) for small applied voltages [9]. It has been shown that piezoelectricity can be well preserved in ultrathin structures which enables a high integration density, miniaturization and reduces the needed voltages for operation [10]. Doubtlessly, the most commonly used piezoelectric material in actuating technologies is $Pb(Zr,Ti)O_3$ (PZT). Nonetheless, $Pb(Mg_{1/3}Nb_{2/3})O_3$-$PbTiO_3$ (PMN-PT) offers unique possibilities in terms of transducing capabilities because of its giant piezoelectric response [11]. This is particularly advantageous for miniaturization considering that the displacement of machined cantilevers made of piezoelectric materials ultimately depends on the square of their length. Recent advances promise good prospects for PMN-PT thin films epitaxially grown on foreign substrates like Si as a platform for MEMS, fully compatible with state-of-the-art silicon technology [9].

As elastic strain engineering (ESE) of materials is attracting increasing interest in the materials science community, advanced actuators capable of exerting enhanced stress fields in materials are needed. The mechanical deformation (strain) in materials – i.e. the change of interatomic distances between their constitutive atoms – affects most of their physical properties. The state of deformation of an element of a material can be described by the strain tensor, that contains six independent components ($\varepsilon_{xx}$, $\varepsilon_{yy}$, $\varepsilon_{zz}$, $\varepsilon_{xy}$, $\varepsilon_{xz}$, $\varepsilon_{yz}$), which are related to the stress components by the compliance tensor of the material (in the linear regime). The deliberate reshaping of the deformation state of a given material is in the heart of ESE, which essentially exploits the tensorial character of the strain field [12]. In this context, piezoelectric actuators are suitable for transferring strain fields to attached films containing embedded active nanomaterials, where the maximum achievable tuning of their properties is eventually conditioned by the elastic limit of the material (~1% in most bulky solid-state structures). Ultra-strength mechanical response has been reported on materials with nanometric dimensions, where the elastic limit scales up as the inverse of a dominant characteristic length related to the size of the material [13]. In particular, nanometer-thick layers (nanomembranes) offer not only increased strength, but they also provide additional functionalities for ESE applications in comparison with bulky structures due to the easy integration, efficient strain transfer and feasible processing by bottom-up and top-down methods [14]. Moreover, in the limit of monolayer thin layers, the so-called two-dimensional (2D) materials feature unprecedented high "stretchability" up to ~20% strain magnitudes, providing a large playground for ESE [12].

Since the revolutionary introduction of uniaxially strained high speed Silicon transistors [15,16] and strained quantum well lasers [17], elastic strain-tunable physical properties of materials have been studied in a wide range of research areas like nanophotonics [18,19], spintronics [20], topological insulators [21] and 2D materials [22]. Controlled strain fields have been introduced in materials by using static approaches like mechanical bending of the sample [23], doping [24], selective etching of pre-strained epitaxial layers [25–27] or thin films deposition on pre-patterned substrates [28]. In such approaches the strain field is irreversibly fixed in the material during the fabrication process. In addition, active materials are often not exempt of unintentional residual strain being introduced during their fabrication/growth process. In an ideal scenario, one could compensate or correct strain-induced effects in the response of the active material by introducing reversible strain fields conveniently employing post-fabrication tuning techniques. Therefore, it is highly desirable to develop tools for reversibly tuning the response of optically active nanomaterials.

In particular, the possibility of tailoring the optical properties of nanomaterials on demand through ESE opens new avenues for both advanced fundamental studies and development of sophisticated devices for future technologies. In this regard, piezoelectric actuators provide a precise and elegant approach to exert reversible strain fields in thin films and nanomembranes integrated on them by simply applying an electrical voltage [18,29]. Another advantage of this approach is that it allows for operation in different environments including cryogenic temperatures and magnetic fields.

The exploitation of the quantum nature of light has triggered an intensive research activity in the field of quantum computation and telecommunications. One major issue that an ideal source of non-classical light should fulfilled is the emission of single photons at a time. Besides single photon emitters like single molecules [30] or nitrogen vacancy centers in diamond [31], self-assembled QDs embedded in a semiconductor matrix have emerged as promising candidates for solid-state applications [32] since they can be electrically addressed upon their deterministic integration in p-i-n diode and/or photonic nanomembrane structures [18]. However, unavoidable structural asymmetries arising from compositional fluctuations in the confining potential barriers for the carriers in the QDs and their shape, as well as the stochastic process in QDs formation, lead to fluctuations in their optical properties which humpers their applicability for advanced quantum optics. Hence, post-processing through the application of external perturbations, i.e. magnetic, electric and/or strain fields, to compensate the inherent asymmetries in self-assembled QDs are mandatory [19,33,34].

On the other hand, semiconductor 2D materials are promising candidates for disruptive multifunctional flexible devices in electronics and optoelectronics [35]. The impressive stretchability of 2D materials may revolutionize the field of ESE due to the unprecedented wide-range tunability of their electronic and optical properties [22]. Very recently, single photon emission in site-controlled predefined positions in semiconductor 2D materials has been demonstrated in monolayer crystals which might open exciting opportunities for quantum optics in 2D materials [36–40]. Most importantly, ESE of 2D materials is still in its infancy and new approaches to study strain effects on their optical properties are highly desirable [22].

In this review, we present recent progress on the investigation and use of strain-tunable optical properties of semiconductors and nanomaterials by employing hybrid devices based on piezoelectric actuators. Section 2 is dedicated to experimental details on sample and device fabrication. The working principles and recent advances in the growth of PMN-PT piezoelectrics are discussed as well as experimental techniques to integrate the active materials on the actuators by bonding, direct transfer and epitaxial growth. Section 3 presents strain-tunable optical emission in nanomaterials by strain fields, employing monolithic piezoelectric actuators, including self-assembled InAs and GaAs QDs embedded in (Al)GaAs nanomembranes and photonic structures, rare earth and metal-ion doped thin films and 2D materials (graphene, $MoS_2$ and $WSe_2$). The tailoring of the optical properties of QDs embedded in stretchable quantum-light-emitting diodes by combining strain and electric fields is shown in section 4. Finally, a novel class of micro-machined piezoelectric actuators for full in-plane stress field control in nanomembranes is shown in section 5. As a demonstration for quantum optics applications, the possibility of generating energy-tunable polarization-entangled photons from self-assembled InAs QDs employing a 6-legged actuator is discussed.

**2.- Device fabrication and working principles.**

**2.1.- Piezoelectric PMN-PT substrates.**

Crystalline materials lacking of inversion symmetry present piezoelectric properties due to the appearance of a net polarization in response to their structural deformation along specific (polar) directions. In the reverse regime, an electrical signal induces the mechanical deformation of the piezoelectric material, which is exploited for actuating applications as discussed in the previous section. Ferroelectric materials, which present a tunable electric polarization (below the Curie temperature) also in absence of a mechanical deformation, feature the highest piezoelectric responses. As mentioned above, the most employed material in MEMS is the ferroelectric PZT, which is mostly used in ceramic form. Achieving large stress magnitudes becomes especially relevant for many applications in miniaturized MEMS, where higher piezoelectric response of the actuating material is highly desirable.

In this regard, the giant piezoelectricity of PMN-PT (a relaxor ferroelectric) represents an ideal alternative with a piezoelectric response about 10 times higher than the PZT counterpart [41]. Such enhanced response is attributed to a phase transition from rhombohedral to tetragonal in PMN-PT (001) oriented substrates when an out-of-plane electric field is applied across the substrate. It should be noted that although (111) oriented PMN-PT substrates offer interesting features such as enhanced anisotropic deformation, they present lower piezoelectric constants than (001) oriented substrates [41]. In addition, there exists a critical $PMN_{(1-x)}$-$PT_{(x)}$ composition for a maximized piezoelectric response, which is given by the morphotropic phase boundary between PMN-rich and PT-rich phases with x=33% [42]. Accordingly, a PMN-PT composition with x~30% is used in most of reported works to ensure a large piezoelectric response while preserving a good compositional stability of the crystal.

For the applications discussed in this review, the typical piezoelectric substrates have a thickness between 200 and 500 μm and they are usually contacted by depositing metallic contacts on both sides of the substrate. Unpoled PMN-PT presents a rhombohedral phase with spontaneous polarization aligned along the [111] crystal direction. The piezoelectric substrate can only be operated after proper poling at room-temperature (RT) by applying an electric field along the [001] direction above a threshold value–typically about 2 kV/cm– at which the polarization rotates towards the [001] direction inducing the transition to a tetragonal phase. Applying an electric field ($\bar{F}_p$) aligned with the poling [001] direction of the crystal produces its out-of-plane deformation (strain $\varepsilon_\perp$) and an in-plane deformation (strain $\varepsilon_\parallel$) as sketched in Figure 1.

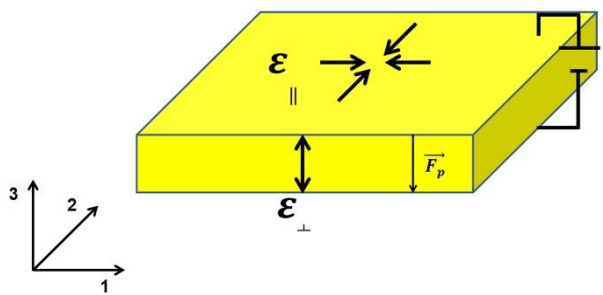

**Figure 1.-** Sketch of a poled piezoelectric substrate where an electric field ($\vec{F}_p$) is applied along the poling direction. The substrate is oriented with the top side being (001). Applying an electric field ($\vec{F}_p$) aligned with the poling [001] direction of the crystal produces its out-of-plane and in-plane deformations (strain $\varepsilon_\perp$ and $\varepsilon_\parallel$, respectively). Highly anisotropic strain fields are obtained at the crystal surfaces parallel to the electric field ($\varepsilon_\parallel \approx -0.7 \times \varepsilon_\perp$). The axis along which the electric field is applied is conventionally labeled with "3", whereas the orthogonal axes "1" and "2" are given to indicate the electromechanical response of the crystal.

The strain can be transferred to active species physically attached onto the piezoelectric substrate. The magnitude and type – compressive/tensile – of the strain can be controlled by simply tuning the amplitude and sign – positive/negative – of the applied electrical voltage, respectively. At the crystal surfaces parallel to the applied electric field, anisotropic strain fields with anisotropy values $\varepsilon_\parallel \approx -0.7 \times \varepsilon_\perp$ can be obtained [43], while compressive in-plane strain magnitudes up to $\varepsilon_\parallel \sim -0.1\%$ can be achieved on top of the substrate at $F_p$=10 kV/cm [44]. The coercive field imposes the threshold for the maximum reverse voltage than can be applied without poling inversion. At low temperatures ~10 K this value is well above 20 kV/cm, which allows applying sufficiently large compressive and tensile stress fields without depoling. In the case of (001) oriented PMN-PT crystals, the in-plane deformation is (theoretically) similar in both orthogonal directions – "1" and "2" in Figure 1– due to a similar value of the piezoelectric constants $e_{31} \approx e_{32}$. (Here, the direction "3" denotes the direction along which the electric field is applied). Other crystal PMN-PT orientations such as (011) present in-plane anisotropy in the piezoelectric constants ($e_{32} > e_{31}$), and offer interesting possibilities to exert highly anisotropic strain fields, when poled along [011] direction, with the major strain axis oriented along the [01-1] direction [45].

Besides approaches based on commercially available substrates, epitaxial growth of PMN-PT thin films on silicon substrates by molecular beam epitaxy, off-axis sputtering or pulsed laser deposition in a process fully compatible with state-of-the-art silicon technology has been carried out by several groups as it would pave the way towards on-chip miniaturized hybrid devices [9,11,46]. There are however many challenges due to complex growth conditions requirements and reproducibility. Extraordinary progress made on high quality PMN-PT epitaxial thin films on silicon substrates has led to impressive piezoelectric constants values as high as $e_{31}$~-27 C/m$^2$ which is about 7 times larger than commercially available bulk substrates [9].

**2.2.- Integration of optically active nanomaterials on PMN-PT substrates.**

In this manuscript, we consider PMN-PT (001) substrates since they are widely used for $d_{31}$ actuating applications, i.e. for exploiting the in-plane deformation (direction "1") for fields applied along the out-of-plane "3" direction. The integration of the optically active nanomaterials/films on the substrates has been carried out by different means as discussed in the following: i) gold thermo-compression bonding; ii) polymer-based "soft" bonding; iii) epitaxial growth; iv) direct transfer by Van der Waals interaction. For approaches i) and ii), the device consists essentially of a nanomembrane with embedded active species that will be bonded on the actuator. In this section we will describe first the experimental procedures commonly used to fabricate such nanomembranes containing QDs or two-dimensional (2D) materials. Finally, the different approaches to transfer the membranes by bonding on the piezoelectric actuators are presented.

**2.2.1.- Nanomembranes fabrication.**

A suitable procedure to introduce strain in optically active nanomaterials – QDs or 2D materials – consists of embedding them in nanomembranes grown/deposited on a carrier substrate. Upon the selective removal of the carrier substrate, these membranes can be then transferred onto the piezoelectric actuator. In this section, we will discuss different procedures in order to obtain nanomembranes consisting of InAs (GaAs) QDs embedded in crystalline matrices, and WSe$_2$ monolayers embedded in amorphous oxides.

Self-assembled semiconductor QDs in different matrices are obtained by epitaxial growth techniques. In the cases presented in this manuscript, the QDs are grown by molecular beam epitaxy (MBE) in a multi-layer structure [GaAs substrate/Al$_{0.7}$Ga$_{0.3}$As sacrificial layer (~ 100 nm)/nanomembrane (thin active layer containing the QDs)]. Two types of QDs will be considered: GaAs and InAs. Specifically, the GaAs QDs are embedded between thin Al$_x$Ga$_{1-x}$As (x<0.5) layers and are obtained by overgrowing nanoholes fabricated by the droplet etching technique [43]. Their typical emission energy ranges from about 1.55 to 1.75 eV. The InAs QDs are embedded in a GaAs host matrix and present optical emission in the spectral range from 1.33 to 1.39 eV. Interestingly, epitaxial growth techniques allow one to have a very precise control (at atomic scales) on the relative position of the QDs layer with respect to top and bottom surfaces. This, together with the possibility of depositing metallic mirrors on the membrane's surfaces, enables designing thin optical cavities where the QDs can be spatially and spectrally coupled to the cavity modes. The thin active layer containing the QDs has typical thicknesses between 150 and 400 nm depending on the final application.

Nanomembranes containing optically active QDs can be released on the host substrate by selective wet chemical underetching. The procedure consists of four steps: i) Au rectangular pads are fabricated on top of the epitaxial multi-layer structure by optical lithography and deposition of a bilayer structure [Cr (5 nm)/Au (100 nm)] by thermal evaporation; ii) non-selective etching with H$_2$SO$_4$:H$_2$O$_2$:H$_2$O (1:8:250) of the whole structure down to the GaAs substrate so that the sacrificial layer is exposed; iii) HF selective etching of the Al$_{0.7}$Ga$_{0.3}$As sacrificial layer [47] releasing micrometer-sized free-standing nanomembranes coated with Au on the substrate. A sketch of the whole procedure is depicted in Figure 2a.

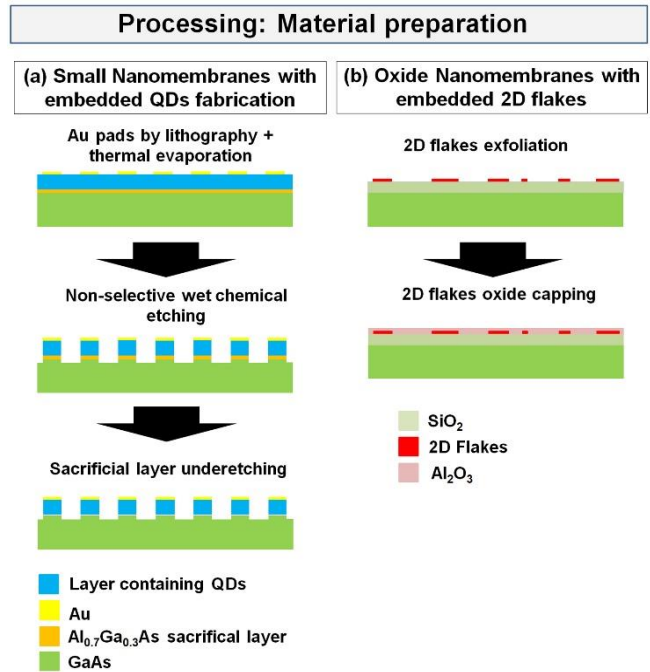

**Figure 2.-** (a) Sketch of the procedure followed to obtain free-standing nanomembranes containing QDs on the host substrate. A sample consisting of a multi-layer structure [GaAs substrate/$Al_{0.7}Ga_{0.3}As$ sacrificial layer (~ 100 nm)/nanomembrane (thin active layer containing the QDs)] is first patterned with Au square pads by optical lithography followed by Cr/Au metallization. These patterns act as masks to define mesa structures using a non-selective wet chemical etching. Finally, the nanomembranes are released on the substrate by performing a selective etching of the sacrificial layer. (b) Sketch of the procedure followed to obtain 2D materials embedded in an oxide nanomembrane structure. 2D flakes are first exfoliated on a 270-nm-thick $SiO_2$ layer deposited on a GaAs(001) host substrate. Afterwards, the flakes are coated with a 30-nm-thick $Al_2O_3$ capping layer.

2D materials' flakes of different thicknesses are conventionally obtained by the so-called mechanical exfoliation technique developed by Novoselov and Geim to isolate graphene monolayers for the first time in 2004 [48]. This is done by employing a scotch tape to thin down bulk crystals in a repeatable process eventually obtaining a dispersed distribution of crystals with different thicknesses on the tape. Afterwards, the tape is brought into contact with a carrier Si substrate and thin flakes are transferred by Van der Waals forces interaction between the flakes and the substrate. The Si substrate must provide good optical contrast for an easy and fast recognition of monolayer flakes under an optical microscope. This is conventionally accomplished by growing thermally a 280-nm-thick $SiO_2$ layer on top of the Si substrate.

For the processing of $WSe_2$ monolayers embedded in nanomembranes, a GaAs substrate can be employed on top of which a 270-nm-thick $SiO_2$ layer is previously deposited by plasma-enhanced chemical vapor deposition. A reference system for flakes localization is used, which consists of gold markers obtained by optical lithography and subsequent Cr/Au deposition. Once the WSe$_2$ monolayers are exfoliated on the substrate and identified with the help of an optical microscope, a 30-nm-thick Al$_2$O$_3$ oxide layer is deposited by atomic layer deposition to encapsulate the flakes. This encapsulation provides protection of the flakes for further processing as described below. The substrate employed in this process has to fulfill two basic requirements: the refractive index should be high enough to provide good visibility to identify monolayer flakes and permit a selective removal in order to release the oxide membrane with the embedded 2D flakes. A sketch of the procedure is depicted in Figure 2b. In principle, any encapsulating oxide can be employed for the nanomembrane as long as it is compatible with the rest of the processing in terms of etching selectivity. We have recently performed a systematic study on the effects of the oxide stoichiometry on the optical emission of the encapsulated flakes that should be taken into account to prevent electrical doping of the flakes and obtain similar emission as as-exfoliated uncapped flakes [49].

**2.2.2.- Nanomembrane transfer by Au thermo-compression bonding.**

Once free-standing membranes are obtained on the host substrate, they are transferred and attached onto the piezoelectric actuators by employing flip-chip bonding techniques. For structures with low height/width ratio the stiffness of the interlayer employed to connect the actuator to the membrane plays a minor role on the strain transfer [50], while high stiffness is desirable for high aspect ratio structures. The Au thermo-compression bonding is a suitable technique since it is well-established for semiconductor wafer bonding applications, and Au possesses a high Young modulus value of about 70 GPa which, in principle, is ideal for an efficient strain transfer. In the bonding process, Au-coated parts to be bonded are brought into intimate contact and pressed together while maintaining a temperature typically around 300 °C for 1-2 hours. Ideally, due to Au inter-diffusion between both parts a uniform stiff bonding is eventually obtained.

To transfer the membranes discussed in the previous section on PMN-PT substrates coated with a [Cr (10 nm)/Au (100 nm)] thin layer, we follow the procedure sketched in Figure 3a. First, the free standing membranes on the GaAs substrate and PMN-PT substrate are brought into contact while applying a pressure of about 10 MPa at a temperature of 300°C during 1 hour. Upon removal of the GaAs substrate, regularly distributed bonded membranes are ideally transferred on the PMN-PT. It should be noted that wrinkles, curling and/or formation of bubbles often observed on underetched membranes can severely degrade the quality of the bonding, which is especially relevant in the case of large membranes with a size of millimeters. To circumvent this problem, a simple process consisting in Au thermo-compression bonding followed by GaAs substrate back-etching allows for an efficient transfer of flat large membranes. In this case, the underetching step discussed above is avoided and the sample is bonded directly on the PMN-PT substrate, thus ensuring intimate contact between both surfaces and, therefore, a better bonding quality. The substrate back-etching is performed in a three steps process (Figure 3b): i) non-selective etching in $H_3PO_4(45\%):H_2O_2(30\%)$ (7:3) to remove most of the substrate and leave a ~50-µm-thick GaAs layer (~45 minutes); ii) selective etching in citric acid (1:1 weight ratio with deionized water):$H_2O_2(30\%)$ (4:1) for the removal of the substrate up to the $Al_{0.7}Ga_{0.3}As$ etch-stop layer; iii) removal of $Al_{0.7}Ga_{0.3}As$ layer with HF(49%) during ~1 minute. (diluted HF or HCl may be used alternatively for this last step).

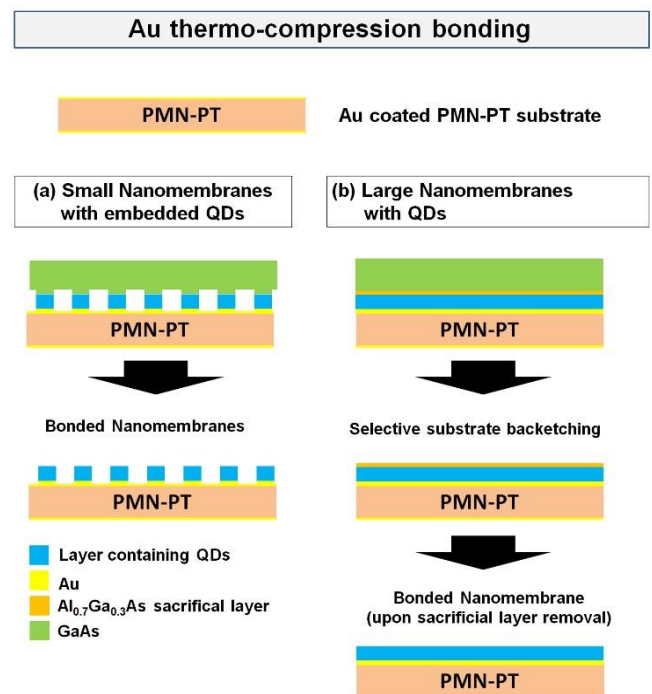

**Figure 3.-** (a) Sketch of the procedure to obtain nanomembranes bonded on PMN-PT piezoelectric actuators by gold-thermocompression bonding. Free-standing nanomembranes are brought into contact with a Au-coated PMN-PT substrate and pressed together (~10 MPa) at a temperature of 300 °C during 1 hour. Bonded nanomembranes are obtained on the PMN-PT actuator upon substrate removal. (b) Sketch of the procedure to obtain large nanomembranes bonded on PMN-PT piezoelectric actuators by gold-thermocompression bonding. A selective removal of the substrate is performed by wet chemical etching followed by etching of the remaining sacrificial layer resulting in a large nanomembrane transferred onto the actuator.

Since the electric field through the piezoelectric actuator is inversely proportional to the substrate thickness for a fixed voltage, it is possible to reduce the working voltages on the actuator by thinning down the piezo to typical values around 300 μm. Unfortunately, mechanical lapping and polishing of PMN-PT material leads to an intrinsic roughness with peak-to-peak values up to ~10 nm due to a different polishing rate of domains with positive and negative polarities [51]. This eventually prevents intimate contact between PMN-PT and sample during the bonding process and the formation of gaps at the interface [18,50]. Some alternatives to circumvent this problem like employing polishing liquids presenting pH factors of 2 has been recently demonstrated [51]. Figure 4c shows a cross-section scanning electron microscopy (SEM) picture of a GaAs nanomembrane with embedded InAs QDs bonded by gold thermo-compression bonding on a piezoelectric PMN-PT substrate as prepared by focused ion beam (FIB) cutting. Clear gaps in the joint layer are visible and attributed to a non-uniform contact between both parts. Alternative approaches to attach nanomaterials to the actuators, as well as comparative studies, are discussed in the following.

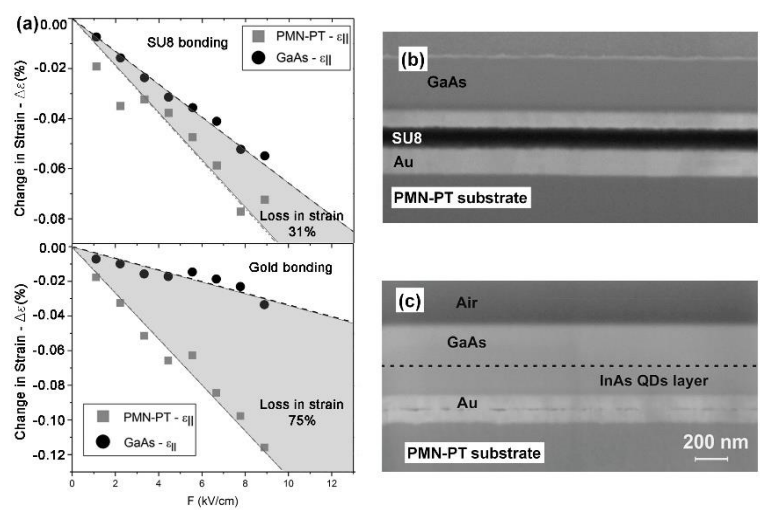

**Figure 4.-** (a) Changes in the in-plane strain component for GaAs bonded nanomembranes and PMN-PT substrate as a function of the electric field applied on the PMN-PT. A comparative study on the strain transfer to GaAs nanomembranes bonded by gold-thermocompression and SU-8 bonding techniques is reported [50]. (b) Cross-section SEM picture of a GaAs nanomembrane bonded on a PMN-PT actuator by SU-8 bonding and (c) gold-thermocompression bonding. A uniform SU-8 bonding interlayer is observed, whereas air gaps can be clearly resolved in the Au bonding interlayer.

**2.2.3.- Nanomembranes transferred by polymer-based bonding.**

Besides a non-uniform bonding, one of the major issues of gold thermo-compression bonding when working with brittle materials like PMN-PT is the relatively high mechanical pressure values required (~10 MPa). In this regard, polymer-based bonding is a good choice since pressures in the kPa regime are usually employed and provides a voidless bonding layer contrary to gold thermo-compression bonding. Among available polymers, Poly(methylmethacrylate) (PMMA), Cyanocrylate and SU-8 have been successfully used for bonding applications [52–54]. The emission energy of the excitonic transitions in self-assembled QDs embedded in semiconductor nanomembranes is very sensitive to the strain field at the QD location and can be therefore used as gauge to monitor the transferred strain. A total energy shift of the exciton transition in InAs QDs embedded in GaAs membranes of about 5 meV has been reported when using Au bonding layers and electric fields varying from -25 to 55 kV/cm [18,55]. Unexpectedly, for similar electric fields (voltages) on the actuator, the same energy shift is obtained in the case where the membranes are bonded to the piezoelectric substrate employing cyanoacrylate [56]. The worse strain transfer was reported when using PMMA as bonding layer (with emission energy shifts of about 1.4 meV), attributed to a lower Young modulus of PMMA, in comparison with the others, combined with the relatively large height/width ratio of the used structures, leading to lateral strain relaxation [57].

SU-8 is a polymer photoresist widely used in MEMS applications that can be patterned by lithography techniques, which also allows a voidless wafer bonding for on-chip applications [54,58,59]. Most importantly, the SU-8 hardens upon UV exposure or thermal treatment at temperatures above the glass transition (~220 °C), leading to Young modulus values up to 5 GPa [58,60]. The latter, together with high optical transparency for wavelengths above about 400 nm make it very promising for optoelectronic applications, where relatively high bonding strengths are required. We will therefore compare the strain transfer capabilities when SU-8 is employed as bonding layer. The SU-8 processing for the devices presented in this manuscript consists in several steps. First, a 500-nm-thick SU-8 layer is spin-coated on the PMN-PT substrate or the sample, depending on the specific application, and soft cured for 5 minutes at 90 °C to evaporate solvents from the layer. The as-prepared coated PMN-PT (or sample) is afterwards brought into contact with the sample (or PMN-PT) and pressed at about 10 kPa while maintaining a temperature of 220 °C for 15 minutes (hard cure of the resist). Figure 4b shows a SEM cross-section image of a GaAs nanomembrane bonded with SU-8 on a gold-coated PMN-PT substrate. Interestingly, a uniform and thin 150-nm-thick bonding interlayer can be observed. A comparative study on the strain transfer from the PMN-PT to the bonded GaAs nanomembrane by Au thermo-compression and SU-8 bonding has been recently reported by performing X-ray diffraction (XRD) studies, where a much better strain transfer efficiency of ~69% is observed on SU-8 bonded nanomembranes with respect to gold thermo-compression bonding (~35%) (Figure 4a) [50]. These findings, which are compatible with those of Ref. [61] for epoxy, demonstrate that polymer-based SU-8 bonding is a more suitable technique in comparison with the others, especially for samples presenting relatively high surface roughness like the case of PMN-PT. The remarkable strain transfer on membranes bonded by SU-8 is attributed to the excellent wettability of the polymer when heated up during the bonding process, which fills the gaps between the PMN-PT and the membrane. Moreover, the observed strain transfer efficiency is higher than that reported on epitaxially grown $SrTiO_3:Ni^{2+}$ or $NdNiO_3/SrTiO_3$ or thin films on PMN-PT substrates [62,63], and similar to epitaxial $La_{0.335}Pr_{0.335}Ca_{0.33}MnO_3$ films [64] with a thickness 10 times lower than the GaAs membranes employed in Ref. [50]. The differences reported above between gold and Cyanocrylate can be attributed to a voidless bonding layer in the latter, similarly to the SU-8 bonding. Figure 5a illustrates the approach used to bond Ga(Al)As nanomembranes containing QDs on PMN-PT substrates by SU-8. Contrary to gold thermo-compression bonding, large membranes with lengths up to centimeters can be easily obtained at a lower temperature and mechanical pressure.

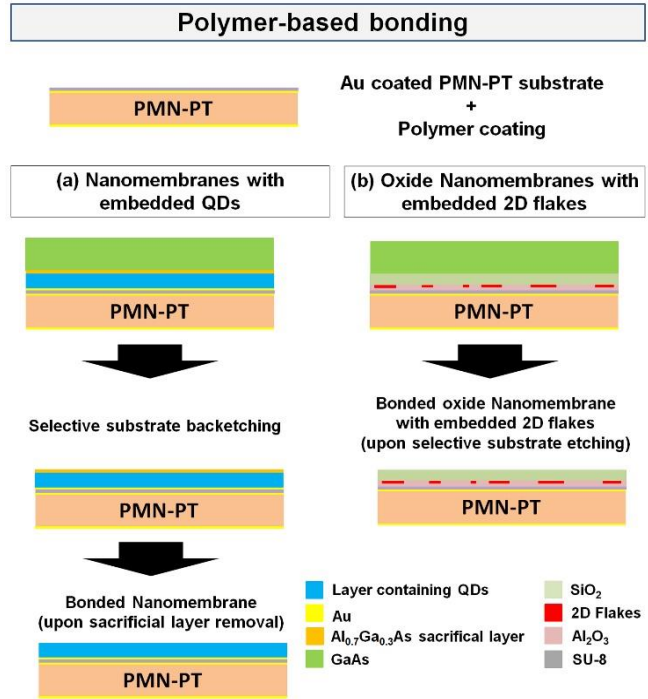

**Figure 5.-** Sketches of the procedure to obtain large nanomembranes bonded on PMN-PT piezoelectric actuators by SU-8 bonding. The sample is first coated with a 500-nm-thick SU-8 layer and brought into contact with the PMN-PT substrate. A pressure of ~10 kPa is applied while keeping a temperature ~220 ºC for bonding of the nanomembrane. Finally, a selective removal of the substrate is performed by wet chemical etching. The case for semiconductor nanomembranes containing QDs and oxide nanomembranes containing 2D flakes are shown in (a) and (b), respectively. A wet chemical etching of the remaining sacrificial layer is performed in the former.

### 2.2.4.- Epitaxial growth.

In principle, the best strategy to maximize the strain transfer is to perform the epitaxial growth of films directly on PMN-PT substrates. SrTiO$_3$ (STO) is known to be a suitable material as a substrate for the epitaxial growth of many complex oxide films. STO has also been used as intermediate layers between PMN-PT (001) and other materials, as it reduces the lattice mismatch between the substrate and the epilayer [63,65]. For instance, modulation of metal-insulator phase transitions in NdNiO$_3$ and magnetization control in La$_{0.7}$A$_{0.3}$MnO$_3$ thin films subject to strain have been successfully demonstrated upon epitaxial growth on PMN-PT substrates [44,63]. In addition, strain engineered tuning of upconversion PL in BaTiO$_3$ films doped with Yb/Er ions and strain-tunable luminescence in STO:Ni thin films have been recently reported [62,66]. Unexpectedly, the strain transfer magnitudes reported in these works do not exceed 46% efficiencies, which is attributed to the semi-coherent character of the interface between PMN-PT and the grown film that allows for strain relaxation mechanisms in the strained film, *i.e.* reduced strain transfer [62,63,65].

Another example of epitaxial growth of $Mn^{2+}$ doped ZnS films on PMN-PT piezoelectric substrates is reported in Ref. [67]. Upon application of dynamical and reversible ac and dc electric fields on the PMN-PT, a strain-induced piezoelectric potential in the ZnS:Mn film is obtained due to the so-called piezo-phototronic effect proposed by Wang [68]. This effect refers to the possibility of tuning the optical properties of materials presenting non-central symmetric structure by a strain-induced piezoelectric potential.

**2.2.5.- Direct transfer of atomically thin 2D materials.**

As mentioned above, the visibility of atomically thin 2D materials on the carrying substrate eventually depends on the optical contrast given by the optical reflectivity of the 2D flake with respect to the top PMN-PT substrate's surface. This is especially relevant for the case of mechanically exfoliated flakes with typical sizes between 5 to 20 micrometers. A suitable procedure consists in growing a transparent oxide layer with the right thickness for maximum visibility in the range of the visible wavelengths around 550 nm. In ref. [69] graphene flakes have been transferred on a 1-µm-thick $SiO_2$ coated PMN-PT substrate, where a thin 60-nm-thick PMMA layer is deposited prior to the direct flake transfer by mechanical exfoliation employing Scotch tape. The PMMA layer is afterwards hardened by annealing at 120 °C to ensure efficient strain transfer from the piezoelectric substrate to the attached 2D flake [69].

Large 2D flakes grown by chemical vapor deposition (CVD) can be directly transferred on top of PMN-PT with no need of any intermediate oxide layer since localization of the flake is not an issue. In Ref. [70], trilayer $MoS_2$ flakes are first grown by CVD on a sapphire substrate and coated by spinning a PMMA layer on top. The coated flakes are then heated at 100 °C for curing the PMMA, and the substrate is selectively etched by using NaOH solution. This process results in a PMMA film with embedded trilayer $MoS_2$ that can be afterwards transferred on the PMN-PT substrate, previously treated with $O_2$ plasma to facilitate the adhesion of the film. The PMMA is removed by acetone, isopropyl alcohol and deionized (DI) water treatment. The top electrode for applying electric fields on the PMN-PT is obtained by simply transferring an optically transparent graphene on top of the structure.

An additional strategy is proposed here to integrate 2D flakes onto PMN-PT actuators. It consists in the attachment of an oxide nanomembrane containing encapsulated 2D flakes on the PMN-PT substrate by SU-8 bonding. Figure 5b shows a sketch of the nanomembrane with embedded 2D flakes on a GaAs (001) substrate attached to a PMN-PT substrate by SU-8 bonding. To release the membrane on the piezoelectric actuator, the GaAs substrate is removed by employing a wet chemical etching with $H_3PO_4(45\%):H_2O_2(30\%)$ (7:3), which is selective against the $SiO_2$ layer.

**3.- Strain-tunable optical properties of semiconductor nanomaterials by means of monolithic actuators.**

PMN-PT piezoelectric actuators have been extensively used in the last years for tailoring the optical properties of various materials including semiconductor QDs embedded in nanomembranes and photonic structures, graphene and semiconductors, rare-earth and metal doped oxide films, as well as piezoelectric ZnS. This section reviews experimental findings on strain-tunable optical properties by employing conventional monolithic actuators. The layout of the monolithic actuator basically consists of a piezoelectric substrate with an optically active material attached to it. The magnitude and anisotropy of the strain fields induced by the actuator when employing a monolithic design can be controlled to some extent by using different piezoelectric crystal orientations and bonding strategies as described in the following.

**3.1.- Strain-tuning of self-assembled quantum dots (QDs).**

Optically active self-assembled semiconductor quantum dots (QDs) are nanostructures with typical dimensions of the order of tens of nanometers. Their fabrication is typically accomplished by epitaxial techniques such as molecular beam epitaxy (MBE). As a consequence of quantum confinement, the allowed energy levels in QDs are discretized similarly to atoms. The most prominent case corresponds to InAs QDs in GaAs matrix, that have been widely explored in the last decades [71]. At sufficiently low temperatures, the photoexcited or electrically injected electrons and holes in the QDs form excitons, which are bound electron-hole complexes. The simplest excitonic species is the so-called neutral exciton (X), an electron-hole bound system. Additional excitonic complexes, such as biexcitons (XX) – two bound electron hole pairs – or trions – composed of two electrons and one hole or two holes and one

electron – can also form depending on the excitation conditions and intentional or unintentional doping. The radiative recombination of XX to the crystal ground state via the intermediate X state produces a cascade of two photons. If the intermediate state is degenerate, this cascade can produced entangled photon pairs, one important resource for quantum technologies [72–75].

Self-assembled QD formation relies on stochastic processes, which lead to fluctuations in the structural and optical properties of the obtained QDs. First, the QD size distribution implies a certain energy distribution of the photons emitted by different QDs. Second, shape and composition fluctuations in the QDs and matrix environment due to alloying with the substrate and capping layer lead to variations of the confining potential for the excitons. Third, residual strain is often present in the self-assembled nanostructures, especially in those obtained via the Stranski-Krastanow (SK) mode. The above mentioned fluctuations have important implications for advanced quantum optics applications where the excitonic emission energy, binding energies (energy difference between X and XX) and energy splitting values have to be precisely controlled [76]. Therefore, post-growth approaches to tailor conveniently the QD optical response are mandatory.

More specifically, the confining potential symmetry lowering from $C_{4v}/D_{2d}$ to $C_{2v}$ in QDs switches on the anisotropic electron-hole exchange interaction, which splits the bright excitons by the so-called fine-structure-splitting (FSS). In real QDs, the symmetry is further lowered to $C_1$ due to strain, alloying and piezoelectricity, as mentioned above. If the FSS is larger than the radiative broadening of the X lines (~1 μeV for typical X lifetimes), the XX has two distinguishable pathways to decay to the ground state, which leads to two transitions with linear polarization [77], spoiling the degree of entanglement of the emitted photons. The probability of finding QDs with FSS<1 μeV is lower than $10^{-2}$ in standard self-assembled QDs, which makes them unpractical for real applications and fundamental experiments [78]. There have been attempts to erase the FSS in QDs by applying a single external perturbation like magnetic [34] or electric fields [79,80]. In the meantime it has become clear that the FSS can be only erased when the perturbation acts along specific directions aligned with the polarization axis of the exciton emission. A lower bound of the FSS is otherwise found due to level anticrossing [33,81,82]. Moreover, in the ideal case of a QD with the polarization aligned along the effective axis of application of the perturbation, erasing of the FSS is achieved at specific X energy. The first demonstration of the universal recovery of the X level degeneray in QDs was reported in Ref. [33] by employing two independent tuning knobs (electric and strain field) with monolithic piezoelectric actuators (see section 4.4). Controlling both, FSS and emission energy, requires the combination of several, independently controlled, "tuning knobs", as discussed in section 5 of this review.

**3.1.1 Control of QD emission energy via piezoelectric-induced strain.**

Strain has a profound effect on the band-structure of semiconductors. In particular, it modifies the energy bandgap of a semiconductor. This effect has been used to control the emission energy of QDs integrated onto piezoelectric actuators. In 2006, Seidl et al. reported on the first demonstration of emission energy and FSS tunability in self-assembled QDs by applying uniaxial stress fields parallel to the QD growth plane [83,84]. In this work, it is successfully demonstrated that uniaxial strain can modify the anisotropies of the exciton wavefunction, and hence the FSS, with negligible decrease of the oscillator strength. In spite this work did not demonstrate suppression of the FSS, it signed the starting point of an intense research activity aiming at investigating the properties of QDs under elastic strains. This approach is based on a PZT piezoelectric stack that allows introducing strain only along a fixed direction so that the strain anisotropy cannot be controlled. Additionally, the relative thick membrane employed in these experiments (0.5 mm) severely limits the magnitude of the exerted stress (~20 MPa). In this regard, nanometer-thick membranes containing self-assembled QDs integrated onto PMN-PT actuators are an ideal platform to efficiently introduce strain fields.

The first work performed with QDs onto PMN-PT substrates was performed by T. Zander et al. [57], and deals with stretching microring structures with embedded QDs and is discussed in section 3.1.2. The tuning of the excitonic optical transitions in QDs may find interesting applications for quantum networking, in particular when the emission energy is tuned to the $D_1$ (or $D_2$) absorption lines of a cloud of rubidium atoms [85]. The main reason is that the atomic cloud can be used as quantum memory for QD photons so as to store and retrieve quantum information, an important ingredient for long distance quantum communication. In the first pioneering work on this hybrid atom-QD interface [83], the authors demonstrate that single photons from GaAs QDs can be slowed down (not stored) when their frequency is finely tuned to the middle of the hyperfine $D_1$ lines of $^{87}$Rb. Energy tuning in this experiment was achieved via magnetic field, an approach requiring bulky equipment and hardly suitable for the envisioned applications. In order to overcome this hurdle, it is possible to integrate high quality GaAs QDs onto PMN-PT actuators and tune their emission energy into resonance with a cloud of $^{87}$Rb atoms by simply using a power supply, as shown in Ref. [56]. Figure 6 shows the energy shift of several excitonic lines stemming from a GaAs QD for a varying electric field applied on the actuator. The total energy shift achieved is ~10.5 meV, a value which is comparable with the inhomogeneous broadening of the QD ensemble. Therefore, it is virtually possible to bring into resonance any arbitrarily selected QD in the sample. It should be noted that the energy shift of the excitonic transitions is mainly due to the strain-induced change of the bandgap of the material, which is proportional to the hydrostatic part of the induced strain field ($\varepsilon_{hyd} = \varepsilon_{xx} + \varepsilon_{yy} + \varepsilon_{zz}$).

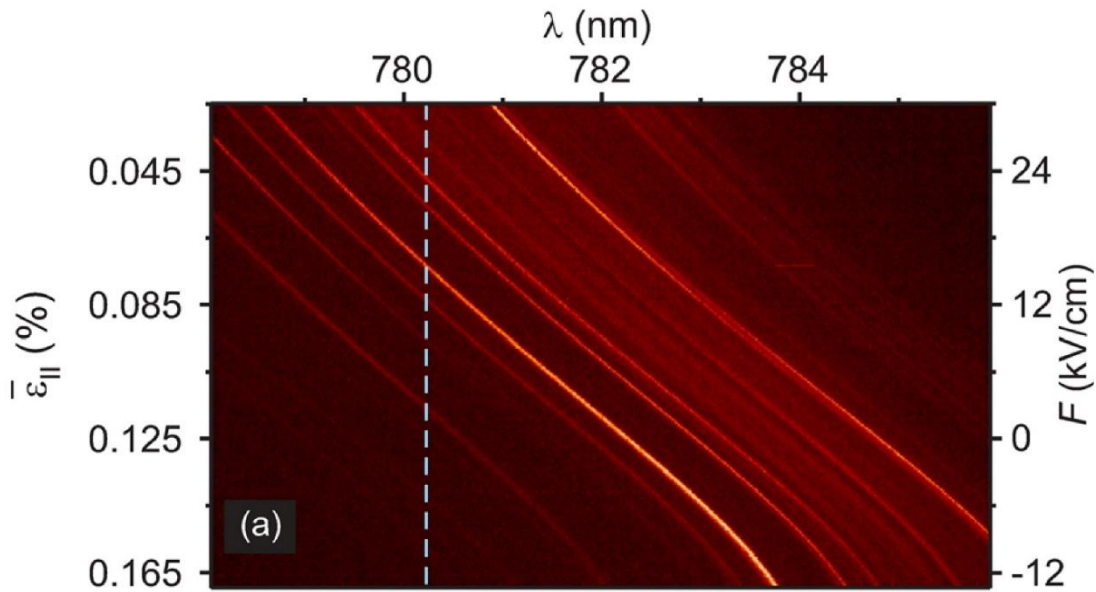

**Figure 6.-** Shift of the exciton energy emission from several GaAs QDs, embedded in a nanomembrane bonded on a monolithic PMN-PT actuator, as a function of the electric field applied on the PMN-PT actuator (right axis) and corresponding in-plane strain (left axis). The total energy shift is as large as ~10.5 meV. Reprinted from [56], with the permission of AIP Publishing.

These results pave the way towards the implementation of hybrid-atomic interconnects with a scalable technology. A further step in this direction has been accomplished in Ref. [86], where the need of an external laser is avoided by injecting carriers into the QD electrically. This is possible by embedding the strain-tunable GaAs QDs into light-emitting diodes (Q-LED) operated in forward bias. The authors demonstrate that it is possible to integrate QD-LEDs onto PMN-PT without degrading the operation of the LED and maintaining the compatibility with the Rb transitions. Recently, the tunability of the excitonic species in QDs integrated on monolithic devices has been successfully exploited in two-photon interference experiments as a first step towards the realization of entanglement swapping between distant QDs [87].

**3.1.2.- Stretchable photonic structures with embedded QDs.**

The efficient extraction of light from QDs by their integration in photonic nanostructures is mandatory for many practical applications involving high yield single photon sources and fundamental quantum-electrodynamics experiments [88–90]. In this section, we will review recent advances on strain-tunable optical properties in optical micro-cavities [57,91,92] and nanowire waveguides upon integration on monolithic PMN-PT actuators [93,94].

In Ref. [93], the fabrication of a nanowire waveguide is performed by etching nanowire structures by reactive ion etching technique on a nanomembrane containing InAs QDs. The membrane is previously bonded on a gold-coated PMN-PT substrate by gold-thermocompression bonding as shown schematically in Figure 7a. In this case, the membrane is released from the substrate by selective etching of a $Al_{0.65}Ga_{0.35}As$ sacrificial layer (see section 2.2 for further details about the processing). The gold layer has a thickness of 200 nm and serves as electrical contact and optical mirror. The QDs are located at a distance D=110 nm from the mirror, which is found to reduce spectral diffusion due to surface states between the nanowire and the Au mirror. These nanowires structures feature single photon emission with large light extraction efficiency values up to 57% for QDs located at the center of the pillar and optimized structural parameters for the pillar geometry. In particular, in these experiments, a nanowire diameter d=223 nm is used for an optimized light extraction at the X emission wavelengths of the ground state in the QDs. Finite element simulations on strained nanowires in Figure 7b show that the strain transfer from the piezoelectric substrate to the pillar decays along the longitudinal direction and the sign is changed from compressive, at the base of the pillar, to tensile due to strain relaxation, which is increased by reducing the pillar diameter. Hence, the magnitude and sign of the strain at the QDs position will eventually depend on their relative position inside the nanowire. Figure 7c depicts the X energy shift when the voltage on the PMN-PT substrate is varied up to corresponding electric field values of ~33 kV/cm for several QDs. The slopes obtained for each QD are clearly different and attributed to different radial positions of the investigated QDs inside the nanowire. This is reasonable considering the large non-uniformities in the strain distribution at different positions in the nanowire where the QDs a located. It is reported a maximum X energy shift of about 1.2 meV that can be reversibly applied with no significant hysteresis.

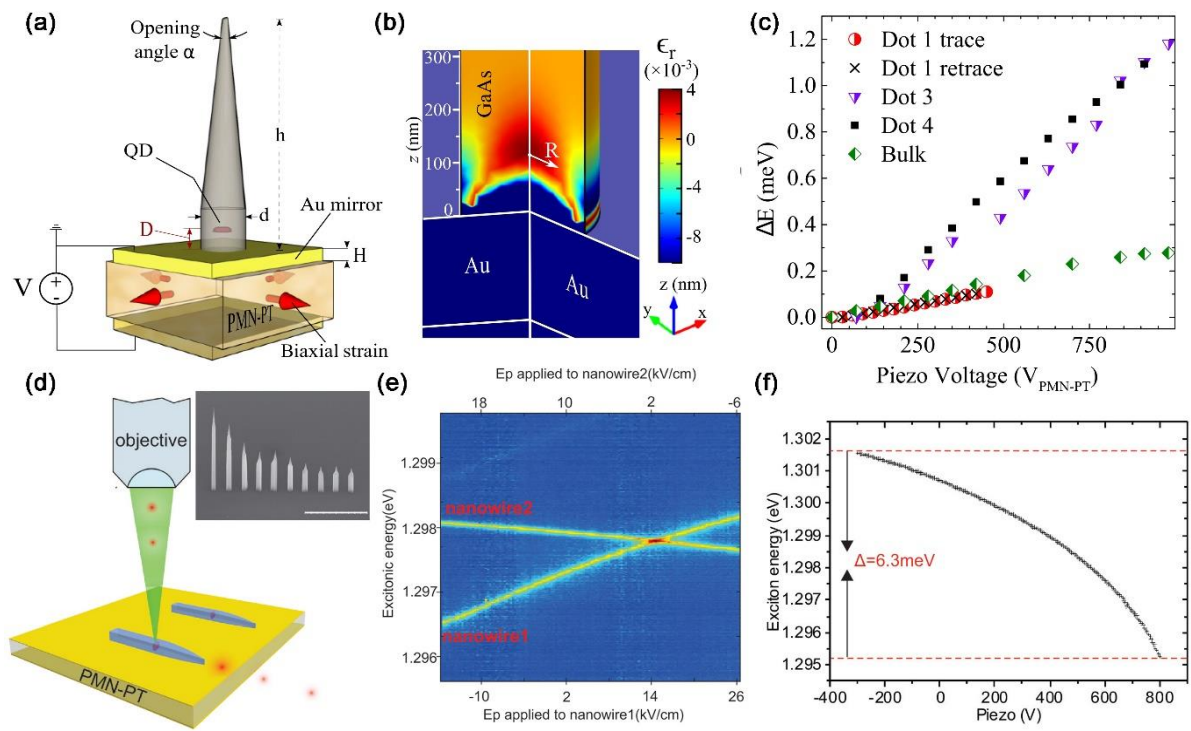

**Figure 7.-** (a) Sketch of an etched nanowire structure on a nanomembrane bonded on a PMN-PT monolithic actuator. The geometrical parameters are optimized to values h=2 μm, D=110 nm, H=200 nm and d=223 nm to enhance optical emission and collection efficiency, while allowing for an efficient strain transfer. Reprinted (figure 1) from [93]. Copyright (2014) by the American Physical Society. (b) Finite element simulation map of the relative strain ($\varepsilon_r = \varepsilon(x,y,z)/|\varepsilon_0|$) distribution in the nanowire, being $\varepsilon_0$ the strain in the PMN-PT crystal. Reprinted (figure 1) from [93]. Copyright (2014) by the American Physical Society. (c) Strain tuning of the exciton X emission as a function of the voltage (electric field) on the PMN-PT actuator for several QDs in nanowires. Reprinted (figure 4) from [93]. Copyright (2014) by the American Physical Society. (d) Sketch of a monolithic PMN-PT actuator where InP nanowires containing InAsP QDs are transfer on top of the actuator. The inset show a SEM image of several nanowires featuring a different tapering value. The scale bar is 3 μm. Reprinted from [94] with the permission of AIP Publishing. (e) Exciton PL emission from two nanowires on different actuators. The emission energies are tuned into resonance by tuning the strain field on the two actuator devices. Reprinted from [94] with the permission of AIP Publishing. (f) Total tuning range for the PL emission from a nanowire for a total electric field on the actuator of ~25 kV/cm. Reprinted from [94] with the permission of AIP Publishing.

Another recent work reports on strain-tunable optical emission from InAsP QDs embedded in InP nanowires obtained by vapor-liquid-solid epitaxy technique [94]. In this case, the position of the QD can be controlled deterministically during growth and the geometrical parameters like tapering angle, diameter and length of the nanowire can be controlled precisely by changing the growth conditions in order to enhance the light extraction efficiency up to 42% [95]. Single photon emission in such nanowires is successfully demonstrated. For their integration onto PMN-PT piezoelectric actuators, a direct-transfer approach with a nano-manipulator is employed (Fig. 7d). Interestingly, the authors report a similar X emission energy tuning range up to ~ 6.3 meV ($F_p$=25 kV/cm) on as-transferred nanowires on the actuator by van der Waals forces interaction and encapsulated nanowires by depositing a structured $Si_3N_4$ oxide layer on top (Fig. 7f). The tuning of the X emission from distant nanowires integrated on different chips is shown in Fig. 7e. Another interesting aspect is that, upon nanowire transfer, the growth axis of the nanowires is parallel to the piezoelectric substrate´s surface which allows in-plane optical emission that might find interesting applications for on-chip integrated devices.

Contrary to the case of nanowire waveguides that allow broadband wavelength operation, in optical micro-cavity structures there are two stringent requirements for an efficient light extraction: spectral and spatial matching of the QD with the cavity mode. The latter can be performed deterministically by in-situ lithography or QD location followed by lithography after growth [96–99]. In these approaches, the position of the QD is known before the fabrication of the cavity structure. Other approaches based on the definition of the QD nucleation sites by ex-situ lithography/patterning techniques with nanometer resolution [100,101], followed by regrowth procedures have also been reported [101]. The spectral tuning of the QD emission with the cavity modes is conventionally realized reversibly by applying magnetic fields [102], electric fields [103] or varying the temperature [104]. However, electric fields lead to PL quenching due to an increased electron and hole wavefunctions separation [105] and temperature tuning is detrimental for the emission efficiency and induce dephasing due to the interaction of the excitons with phonons.

Strain-tuning to frequency-match the QD photons with the cavities modes by employing piezoelectric actuators has been demonstrated as a suitable strategy to overcome these problems in an elegant manner. The key ingredient is related to the fact that the QD and the cavity mode shift at a different rate when stress is applied.

Strained QDs inside micro-ring optical cavities have been reported by integration on PMN-PT actuators in Ref. [57]. The sample consists of an epitaxial 250-nm-thick GaAs layer – containing InAs QDs located at the center – on a $Al_{0.7}Ga_{0.3}As$ sacrificial layer grown by MBE on a GaAs(001) substrate. The micro-ring structures are first fabricated by electron-beam lithography leading to patterned features replicating the ring-like shape, which is used as a mask to etch the material around the structures down to the substrate. Figure 8a shows a SEM picture of the etched micro-ring where the GaAs layer with embedded QDs and the $Al_{0.7}Ga_{0.3}As$ sacrificial layer are well distinguished. The micro-rings are released from the substrate by underetching the sacrificial layer with HF (Figure 8b) and transferred onto the PMN-PT actuator by PMMA bonding. A 1-µm-thick $SiO_x$ layer is deposited on the PMN-PT by thermal evaporation to prevent light absorption losses. Finally, the bonded microrings are coated with a thin PMMA layer to help for an efficient strain transfer. An optical image with an array of microrings bonded on the PMN-PT substrate is shown in Figure 8c.

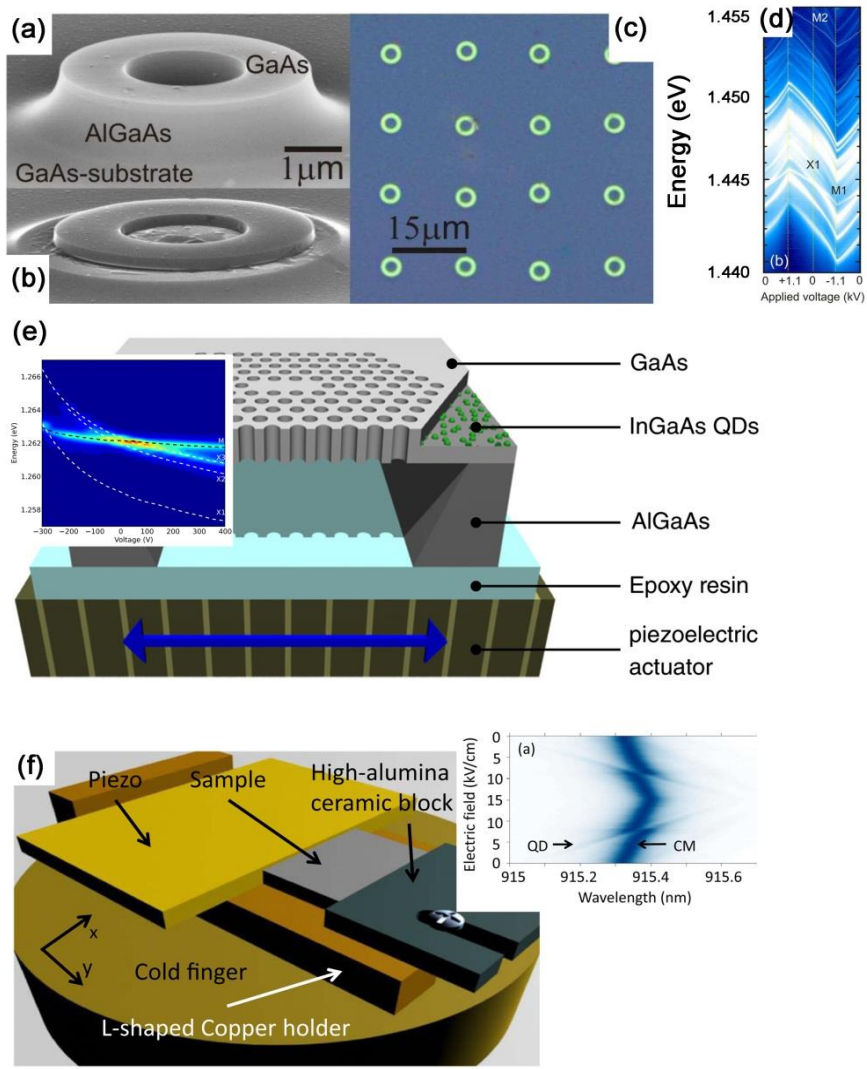

**Figure 8.-** (a) SEM picture of a micro-ring structure etched on a GaAs layer with embedded InAs QDs at the center and a $Al_{0.7}Ga_{0.3}As$ sacrificial layer epitaxially grown on a GaAs(001) substrate by MBE. From Ref. [57]. (b) SEM picture of the structure upon sacrificial layer selective removal by HF, where free standing micro-ring structures are eventually obtained on the substrate. Reprinted from [57]. (c) Optical picture of bonded micro-rings on a PMN-PT monolithic actuator. Reprinted from [57]. (d) excitonic X1 emission energy and cavity modes M1/M2 shifts as a function of the applied voltage on the actuator. Reprinted from [57]. (e) Sketch of a suspended L3 photonic crystal cavity (embedding QDs) bonded on a PZT piezoelectric stack actuator that allows introducing uniaxial strain fields along the defect line of the cavity. The inset shows the excitonic X1, X2 and X3 emission from different QDs which can be successfully tuned to the cavity mode M by applying a varying voltage on the actuator. Reprinted from [91]. (f) Sketch of a sample consisting of a photonic crystal with embedded QDs installed in a L-shaped copper holder coupled to a PMN-PT piezo actuator. The inset shows the excitonic X emission of a QD and cavity mode (CM) shifts as a function of the electric field applied on the actuator. Anticrossing of both X and CM reveals QD-cavity resonance in the strong coupling regime. Reprinted from [92] with the permission of AIP Publishing.

These micro-rings cavities present a quality factor exceeding 12000. Interestingly, the X emission energy ($E_x$) and optical mode energy ($E_M$) shift almost linearly with the applied voltage on the PMN-PT with a total shift value of about ±3.8 meV for $E_x$ ($F_p$~74 kV/cm) as shown in Figure 8d. The relatively low strain values with respect to other reported works are attributed to a much softer PMMA bonding layer that does not allow an efficient strain transfer. Most interestingly, $E_x$ and $E_M$ shift in the same direction but with different rates which means that the matching condition $E_x = E_M$ can be reached in order to tune into resonance the QDs emission with the cavity mode. No PL emission linewidth broadening or quenching, usually observed when using electric fields or temperature to tune the optical emission, are observed. As expected, there is some dispersion in the energy shift values for different QDs associated with non-uniformities in the strain field distribution, similarly to the results from the etched nanowire structures shown above. The change of $E_x$ with strain is attributed to the bandgap tuning of the material, being larger (smaller) for compressive (tensile) strain fields, whereas the change of the cavity modes under tension is attributed to both increased size of the resonator and increase of the refractive index because of the photoelastic effect [57]. Similar experiments based on microdisks and reaching also the strong coupling regime were reported in Ref. [106].

Photonic crystal optical cavities with embedded QDs have been demonstrated as an efficient strategy for enhancing light extraction. Such a structure essentially consists of a defect in a periodic arrangement of holes fabricated by e-beam lithography followed by reactive ion etching around a single QD in a suspended semiconductor membrane. The design of the holes size and arrangement as well as the thickness of the membrane is selected to couple the optical emission of the QD with the cavity mode. In Ref. [91], a L3 photonic crystal cavity is fabricated (quality factor Q~3310) on a sample grown by MBE on a GaAs(001) substrate that consists of a GaAs nanomembrane with InAs QDs at its center and a $Al_{1-x}Ga_xAs$ sacrificial layer with high Al content (x~0.7). The GaAs substrate is selectively removed by combining mechanical polishing and wet chemical etching. The resulting slab is afterwards bonded on a PZT piezo stack actuator by using an epoxy resin and the photonic crystal structure is underetched by using HF. The defect line of the photonic crystal is aligned along the main stretching direction of the actuator that allows exerting almost uniaxial strain fields. A sketch of the final device is depicted in Figure 8e. Upon application of a voltage on the piezo actuator from +400 to -300 V (strain variation up to 0.08%) the QD emission lines shifts in energy 5 times faster than the cavity mode. This is relevant, as it allows one to couple the QD emission to the cavity mode as reported in the inset to Figure 8e.

Another approach to introduce strain in photonic crystals is reported in Ref. [92]. In this case, the sample containing the photonic crystal (quality factor Q~12000) is installed on a device that consists of a L-shaped copper holder coupled to a PMN-PT piezoelectric substrate as shown in Figure 8f. This configuration allows applying uniaxial stress fields. Similar to the previous case, the row defect of the cavity is aligned along the direction of the applied uniaxial stress. The QD exciton emission and cavity mode shift as a function of the electric field applied on the PMN-PT up to 15 kV/cm, as shown in the inset to Figure 8f. An anticrossing of the cavity mode and QD emission is observed at a field 7.8 kV/cm, a clear indication of a QD-cavity coupling in the strong coupling regime. Unexpectedly, a redshift instead a blueshift is observed on the QDs emission under compressive strain, a fact probably arising from curling or wrinkling of the suspended structure. It should be noted that also fluctuations in the In content of the QDs may give rise to blue-shifts of the emission under compression, see Ref. [107]. Due to a non-uniform distribution of the applied stress field, a different tuning range is observed on various QDs. In line with the works mentioned above, the average shift of the QD emission is higher than the cavity mode with a QD to cavity mode tuning ratio of about 5.8.

**3.1.3.- Tuning the FSS in QDs.**

The tuning of the FSS with PMN-PT actuators was reported later on in Ref. [43], where excitons confined in quantum-well potential-fluctuations (QWPFs) are studied. In this experiment the nanomembrane is bonded on the side of a PMN-PT (001) substrate, which permits delivering highly anisotropic strain fields $\varepsilon_\perp \approx -0.7 \times \varepsilon$ (see Figure 1) that are more suitable to modify the anisotropy of the QD confining potential. Figure 9a shows the PL polarization-resolved maps for the exciton emission from a QWPF where the splitting of the X bright states (FSS) is well resolved. In addition to the inherent anisotropies in the confining potential in the as-grown QDs, the fabrication process employed to transfer the membrane onto the piezoelectric actuator increases the anisotropy of the strain field in the QD. While continuously varying the applied electric field in the piezoelectric actuator from 33 to -20 kV/cm, the two split X states undergo a clear anti-crossing accompanied by a rotation of the polarization direction of the exciton emission, which mainly occurs when the FSS reaches its minimum value (see Figure 9b). Most importantly, the polarization angle saturates for a large strain magnitude, which roughly corresponds to the major strain axis direction of the strain field (strain direction). Based on a continuum model, the observed effects can be attributed to substantial changes of the hole states which eventually condition the polarization of the emission. We refer the reader to Ref. [43] for further details. The same trend is observed in the case of GaAs QDs and InAs QDs, as reported by the authors within the same work. A maximum FSS tunability for GaAs QDs and InGaAs QDs of ~70 µeV and 25 µeV are found, respectively. It should be noted that in this experiment the anisotropy and direction of the strain field are fixed and only the magnitude of the applied stress is varied. In turn, this implies that X bright states generally undergo anti-crossing vs the applied voltage, as the major strain axis is not properly aligned to fully compensate for the QD structural asymmetries.

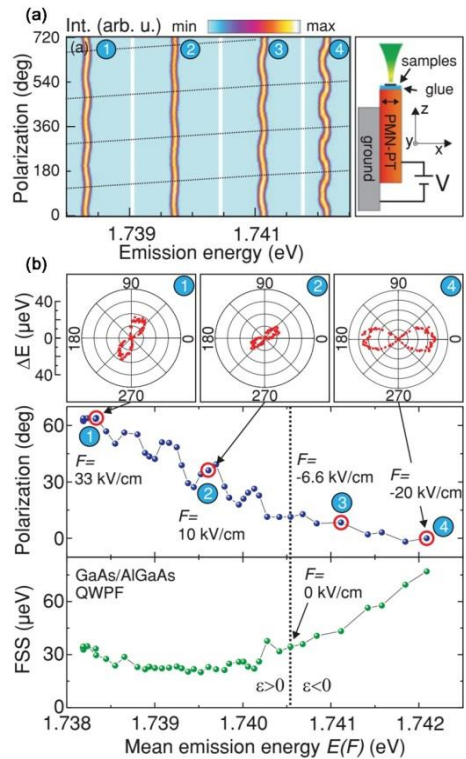

**Figure 9.-** (a) Polarization-resolved maps for the emission of an exciton confined in a QWPF in a GaAs/AlGaAs quantum well. The observed FSS changes for different exerted strain fields (electric fields in PMN-PT actuator). (b) Evolution of the polarization angle and FSS as a function of the mean emission energy obtained for different values of the electric field across the PMN-PT actuator. Reprinted (figure 1) from [43]. Copyright (2011) by the American Physical Society.

**3.1.4.- Interfacing photons with Cs atomic vapors by straining QDs.**

Monolithic piezoelectric actuators can be employed to tune the optical emission from InAs QDs through the two absorption lines of $D_1$ transitions in a Cs vapor of atoms, which has interesting applications as a spectrally selective delay line for single photons [108]. Figure 10a shows a sketch of the Cs $D_1$ transitions that are split into the $6\ ^2S_{1/2}$ and $6\ ^2P_{1/2}$ levels. These levels are further split due to hyperfine coupling into levels with total atomic angular momentum F=3 and F=4. The optical transmission around the $D_1$ lines reveals the absorption due to the characteristic four transitions of the hyperfine structure for different vapor temperatures ($T_{Cs}$), as shown in Figure 10b. Only two absorption dips separated by 10 GHz are resolved for $T_{Cs}$>100 °C due to Doppler broadening. The splitting of the $D_1$ lines is significantly larger than the one observed for $^{87}$Rb atoms [86]. This is advantageous when working with QDs featuring relatively large bandwidth. Moreover, if the FSS value of a given QD is larger than the splitting between the two absorption Cs $D_1$ lines (41 µeV), it is possible to delay selectively each of the orthogonally polarized split components, horizontal (H) and vertical (V), for the X or XX lines, which would open the possibility of realizing time reordering of both lines. We refer the reader to Ref. [108] for further information. Figure 10c shows the X split bright states in a QD with a FSS~58.7 µeV. The corresponding PL amplitude when each component is tuned through the Cs $D_1$, upon the application an electric field ($F_d$) on a diode structure, is reported in Figure 10d. When one of the components is tuned to Cs $D_1$, it can be selectively delayed with respect to the other, as shown in the time-resolved measurements in Figure 10e-h. The time delay depends on the Cs atomic vapor temperature, being larger as the temperature is increased. The delay times reachable for a vapor temperature of 143 °C can be as high as 4.9 ns. The observed temporal distribution of the arriving photons is also modified and can be fully explained by considering the spectral broadening of the QD emission and the atomic lines.

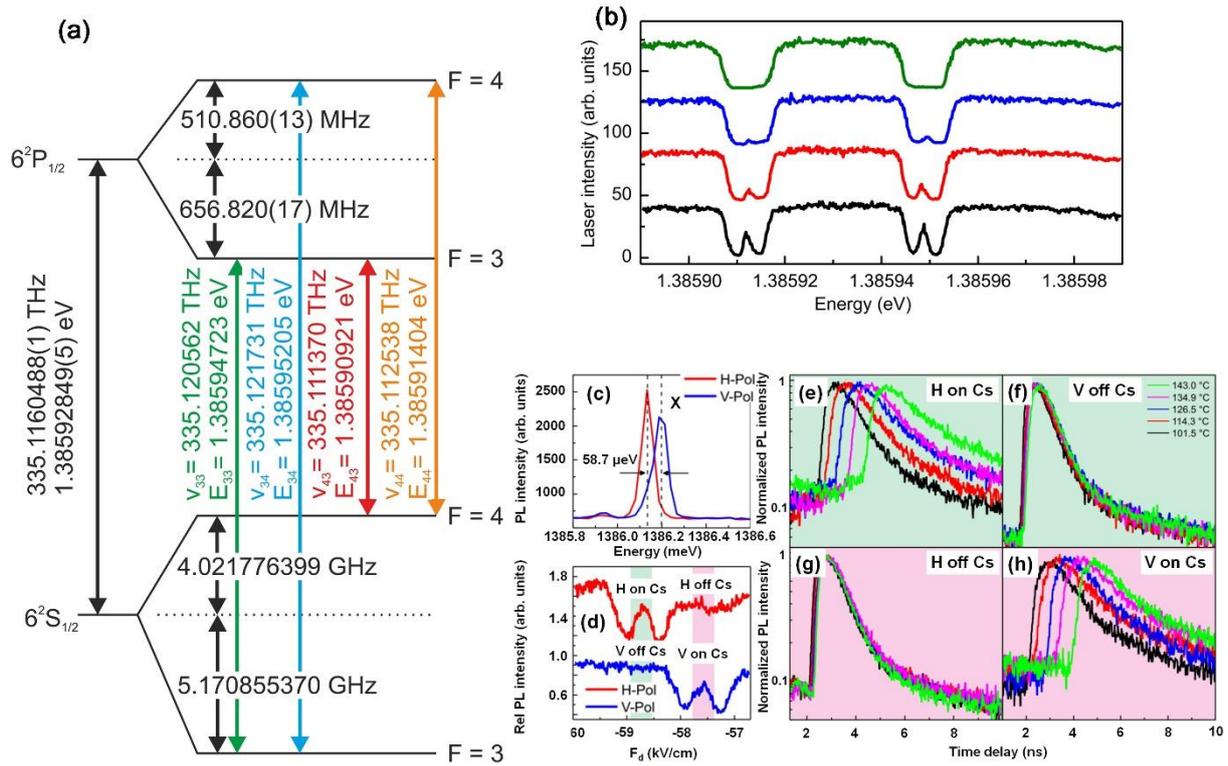

**Figure 10.-** (a) Split $6\,^2S_{1/2}$ and $6\,^2P_{1/2}$ levels of the $D_1$ transitions in a vapor of Cs atoms. Due to hyperfine coupling, these transitions are further split into levels with total atomic angular momentum F=3 and F=4. (b) Optical transmission spectra around the $D_1$ levels for several Cs vapor atoms temperatures. The four characteristic transitions of the hyperfine structure are revealed through absorption features. (c) Split exciton X bright states with orthogonally polarized components H and V. (d) PL intensity for each H and V components as they are tuned through the $D_1$ Cs transitions by varying the electric field applied on the diode structure ($F_d$). (e-h) Time resolved PL measurements of the X bright states for several Cs atoms vapor temperatures when components H and V are on resonance (on) and out of resonance (off) with the $D_1$ transitions. Reprinted (figure 1 and figure 3) from [108]. Copyright (2015) by the American Physical Society.

### 3.2.- Straining 2D materials.

2D materials have been largely explored since the first isolation of graphene in 2004. Among them, semiconductor 2D transition metal dichalcogenide (TMDCs) materials are very interesting for optoelectronic applications since they present relatively large in-plane mobility (up to 200 cm$^2$ V$^{-1}$ s$^{-1}$), strong light-matter interaction and strong exciton-binding energies up to 1 eV at room temperature (tens of meV for conventional semiconductors), which makes them ideal candidates for optoelectronic, photonic and electronic applications [35]. Most importantly, TMDCs present a sizeable band-gap that ultimately depends on the material thickness (number of layers) and/or its state of deformation (strain) [22,109]. Contrary to conventional semiconductors like group-IV, III-V or II-VI compounds when scaled down to atomic thicknesses, the surface of 2D materials presents no dangling bonds and therefore, an enhanced device performance is naturally expected. Recently, single photon emission from localized centers in few-layers $WSe_2$ and hBN 2D materials has been reported as promising candidates for applications in quantum optics [37,110]. In the context of this review, one of the major advantages of 2D materials compared to conventional semiconductors is their high stretchability up to values around 20% without plastic deformation, which offers a large playground for elastic strain engineering of their electronic and optical properties. This is especially interesting in 2D semiconductors, since strain provides a natural strategy to tune the band-gap of the material.

There have been many approaches to introduce strain fields in these materials, in particular using substrate bending, where deformation is induced mechanically or thermally. In addition, flake wrinkling on flexible substrates and substrate deformation by heating with a laser have also been demonstrated [22]. Despite all the progress, the field of strain-engineering of 2D materials would benefit from strategies to introduce the strain fields with magnitude and sign, which could be controlled precisely in a reversible way. This field is still in its infancy and offers interesting opportunities to explore new physics in such systems.

In Ref. [70], the integration of CVD grown $MoS_2$ tri-layer flakes on piezoelectric PMN-PT monolithic actuators by direct transfer has been successfully demonstrated (Figure 11a). Such approach benefits from a higher strain transfer, as no intermediate oxide layer is needed to enhance the flake visibility due to the large lengths of the obtained 2D flakes. Reversible biaxial compressive strain fields up to 0.2% are reported with a total energy blue shift of about 60 meV for the direct band-gap PL transition (Figure 11b). This result demonstrates that the Van der Waals forces between the flake and bare PMN-PT substrate surface are strong enough to allow an efficient strain transfer, an important result as it proves that bonding layers can be avoided in these systems. Moreover, since the top Au electrode used to contact the piezoelectric actuator is replaced by a graphene layer transferred on top of the $MoS_2$ flake (see the Figure 11b), a more efficient strain transfer is in principle possible. The PL emission shift is accompanied by an intensity enhancement up to 200% and FWHM reduction by about 40%. The authors attribute these effects to an increase of the density of states of the carriers due to strain-induced band structure modification in MoS$_2$.

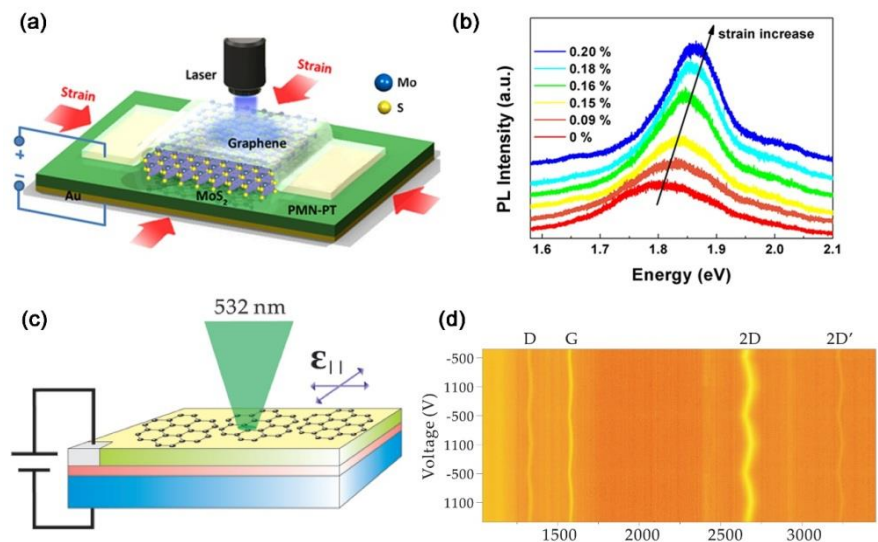

**Figure 11.-** (a) Sketch of a PMN-PT monolithic device where a tri-layer MoS$_2$ flake is transferred on top. A top graphene layer is employed as top contact for the actuator. The bottom side of the piezoelectric is coated with Au and used as electrical ground. Adapted with permission from [70]. Copyright (2013) American Chemical Society. (b) PL spectra of the MoS$_2$ flake as a function of the electric field applied on the actuator. A clear PL blueshift is observed for induced compressive strain. Adapted with permission from [70]. Copyright (2013) American Chemical Society. (c) Sketch of a PMN-PT monolithic device where a graphene monolayer flake has been transferred. In this case, the piezoelectric substrate (blue) is coated on top with a 40-nm-thick LSMO layer (red) and at the bottom with an Au layer for electrical contacts. A 1-µm-thick SiO$_2$ (green) is deposited before transferring the flake. A PMMA layer (yellow) is spinned on top for an efficient flake transfer. Adapted with permission from [69]. Copyright (2010) American Chemical Society. (d) Raman spectroscopy map as a function of the applied voltage to the actuator. Adapted with permission from [69]. Copyright (2010) American Chemical Society.

Graphene monolayers have been also integrated on PMN-PT actuators in Ref. [69]. In this case, the piezoelectric substrate is coated on one side with an Au layer and on the other side with 40-nm-thick epitaxial layer of La$_{0.7}$Sr$_{0.3}$MnO$_3$ (LSMO) to ensure electrical contacts. On top of this layer, a 1-µm-thick SiO$_2$ followed by a 60-nm-thick PMMA layer are deposited as strain-transfer layer for the flake (Figure 11c). The characteristic Raman features D, G, 1D and 2D' experience a clear linear and reversible shift towards lower/higher frequencies for in-plane tensile/compressive strain fields introduced upon application of a voltage to the piezoelectric substrate (Figure 11d). No flake wrinkling or slippage is found after several cycles, that indicate a good mechanical stability of the flake. The maximum achievable strain magnitude is around 0.15% in this case. The different strain values with respect to the previous work might be related to the interlayers between the flake and the piezoelectric that may influence the efficiency of the strain transfer.

We present now a new approach to introduce strain in WSe$_2$ monolayers by embedding the flakes in an optically transparent oxide nanomembrane that is bonded onto a PMN-PT monolithic actuator. The piezoelectric substrate is coated with a 100-nm-thick Au layer on both sides for electrical contacts. The fabrication of the oxide nanomembranes containing the flakes consists of several steps. First, WSe$_2$ flakes are produced by mechanical exfoliation on GaAs (001) substrates coated with a 280-nm-thick SiO$_2$ layer to ensure visibility of monolayer flakes under an optical microscope. The selected flakes are localized with respect to Au markers defined on the SiO$_2$ surface by conventional optical lithography and metal evaporation. Afterwards, the flakes are encapsulated with an amorphous 30-nm-thick Al$_2$O$_3$ capping layer deposited by atomic layer deposition to protect them during further processing. We point out that flakes encapsulation in oxide nanomembranes is a suitable strategy to avoid wrinkling or slippage of the flakes [49]. The coated GaAs substrate containing the localized encapsulated flakes sample is bonded on the actuator by using SU-8 bonding technique (see section 2.3.3.). The bonded oxide nanomembrane is then released on the PMN-PT upon GaAs substrate selective removal by wet chemical etching, as described in detail in section 2.2 of this manuscript. The PMN-PT with the bonded nanomembrane is finally electrically contacted with silver glue on an AlN chip carrier, which provides good electrical insulation and thermal contact for low temperature studies. The sketch and optical pictures of the final device are shown in Figure 12.

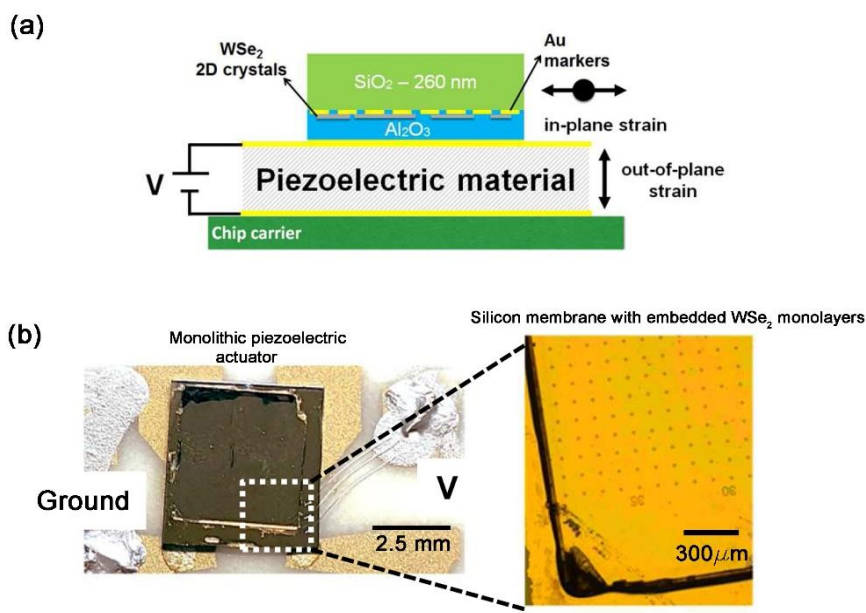

**Figure 12.-** (a) Sketch of a PMN-PT monolithic device where an oxide nanomembrane containing WSe$_2$ monolayer flakes is bonded by SU-8 bonding technique. (b) Optical image of the device contacted on a home-made AlN chip carrier. A detail of the optically transparent nanomembrane is shown, where markers for flakes localization are visible.

A micro-PL spectrum corresponding to the direct band-gap optical emission in a WSe$_2$ monolayer at RT is shown in Figure 13a, where a prominent peak is resolved at around 1.655 eV for an excitation power of 10 µW using a continuous-wave laser emitting at 532 nm. The laser is focused on the sample with a 50x objective with a spot size of about 1 µm. The optical emission is shifted reversibly by sweeping the voltage applied to the PMN-PT towards positive values, which introduce a compressive in-plane strain field. A total blueshift of about ~4 meV is found for electric fields up to $F_p$=6.6 kV/cm, a feature that is attributed to the change of the WSe$_2$ monolayer band-gap (Figure 13b). Based on recent comparative studies by XRD to estimate the strain transfer efficiency [50], an in-plane compressive strain field of about 0.05% is estimated on the WSe$_2$ monolayer, which corresponds to a band-gap tuning of about 80 meV/%, a value which is comparable to other reported values in semiconductors [111]. However, we note that the observed value is much lower than the 300 meV/% reported on MoS$_2$ trilayers directly transferred on the PMN-PT actuator [70]. This might be due to a better strain transfer efficiency for the work reported in [70], since the flake is in direct contact by Van der Waals forces with the piezoelectric substrate. In our case, the strain is transferred through the 100-nm-thick Au layer deposited on top of the piezoelectric, the ~200-nm-thick SU-8 bonding layer and oxide nanomembrane where the flakes are embedded.

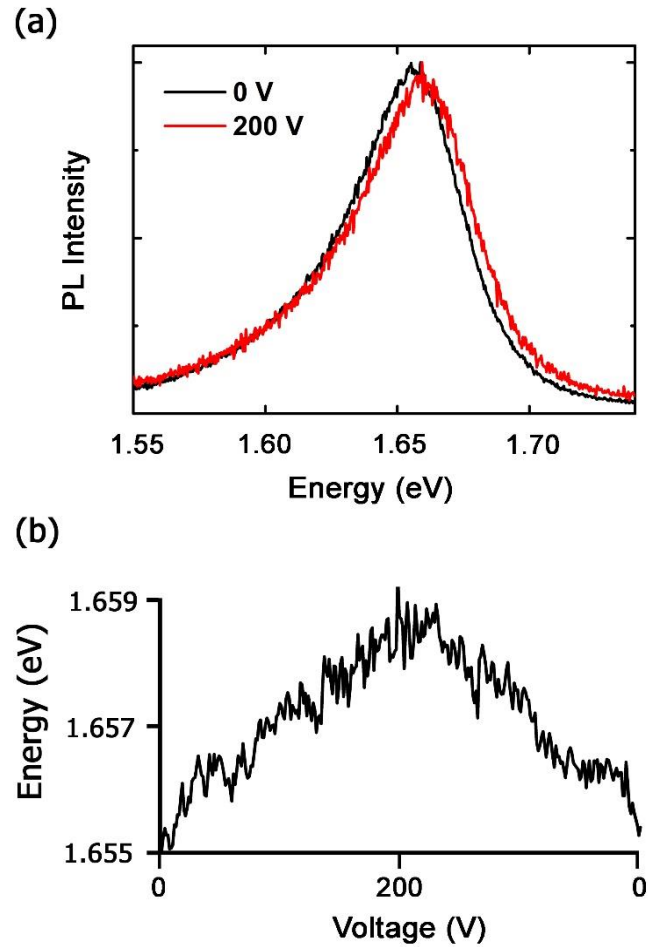

**Figure 13.-** (a) Direct band-gap micro-PL spectrum of a WSe$_2$ monolayer encapsulated in an oxide nanomembrane. (b) PL peak position shift as a function of the voltage applied on the PMN-PT piezoelectric actuator. A total shift of ~4 meV is observed for equivalent electric fields of ~6.6 kV/cm.

As mentioned above, one of the main advantages of this kind of devices is that they can be installed in vacuum conditions inside a cryostat for low temperature measurements. The PL emission for the WSe$_2$ monolayer at 10 K presents several features related with neutral and charged excitonic complexes [49,112]. Besides these transitions, PL emissions with linewidth values ~ 1 meV and below can be resolved in some flakes, associated with the radiative recombination of localized excitons. The origin of such transitions is still unclear. They have been attributed to either presence of defects in the monolayers or strain-induced split impurity states due to the wrinkling of the flake [36,39]. Interestingly, these transitions have demonstrated to deliver single photons [36,37,113], a feature that may open new routes for the exploitation of 2D materials for quantum optics applications. The PL spectrum of a WSe$_2$ monolayer flake presenting one of such sharp transitions (linewidth ~ 1.8 meV) is shown in the inset of Figure 14a. Remarkably, the PL emission energy can be shifted by introducing a strain field with the PMN-PT actuator, as shown in Figure 14a. In this case, the emission energy was tuned around 3 meV for a total field $F_p$=16 kV/cm (compressive strain). The different tuning range observed at RT and 10 K is due to the lower (roughly 25%) piezoelectric response of the PMN-PT at low temperatures [114]. This result is in line with other reported works which relate the origin of such emissions to the strain state of the flake [39,40]. By encapsulating the flakes, reproducible energy shifts and stable emission are found in flakes encapsulated in nanomembranes. We therefore conclude that oxide encapsulation of the flakes is a suitable strategy for the integration for 2D materials on piezoelectric actuators. We refer the reader to Ref. [49] for further additional information about oxide encapsulation effects on the optical properties of $WSe_2$ monolayers.

To further investigate the PL emission related to localized excitons, we measured polarization-resolved PL at 0 V, as shown in Figure 14b (left panel). Two different components with nearly orthogonal polarization can be resolved with different intensities. These features have been previously reported and associated with a FSS [37]. The PL peak was fitted with two Lorentzian curves against the polarization angle of the PL signal and a value FSS~500 µeV was estimated when no voltage is applied on the piezo. Interestingly, upon application of 500 V on the PMN-PT, a FSS increase up to ~640 µeV is observed together with the PL shift. This experimental observation might suggest that the nature of such transitions in monolayer flakes is related with the deformation state of the flake, in agreement with recent works where static strain fields are introduced by means of pillar stressors [39,115]. Furthermore, it is potentially possible to tailor their emission energy as well as the anisotropies associated with the confining potential by elastic strain engineering through more sophisticated actuators [29].

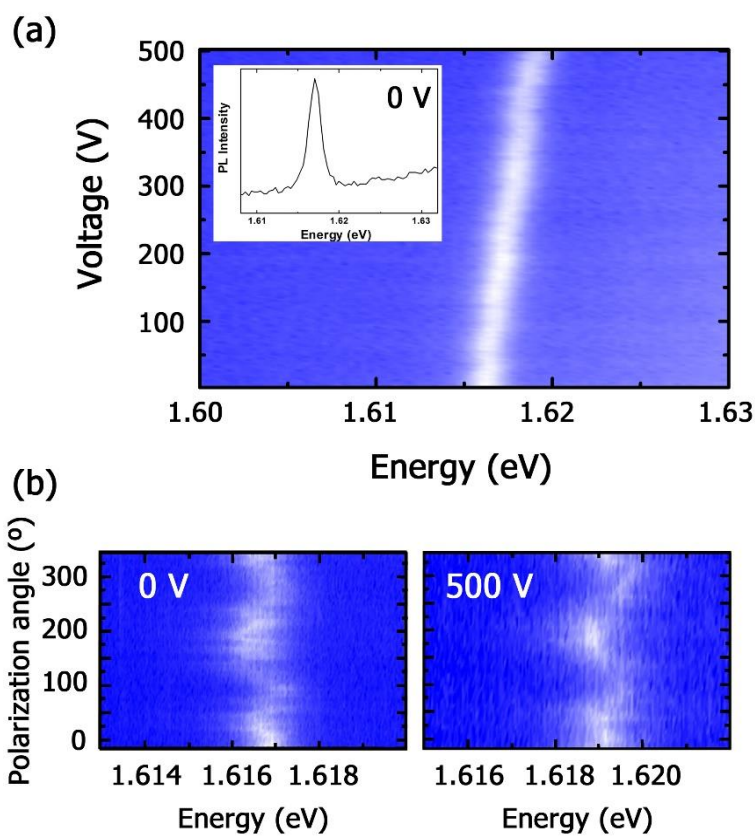

**Figure 14.-** (a) Color-code PL spectra from a WSe$_2$ monolayer as a function of the applied voltage on the PMN-PT actuator. The inset shows the sharp PL signal associated to the radiative recombination of a localized exciton in the flake. A clear blueshift of the PL signal of ~3 meV is resolved when the voltage is swept. (b) Polarization resolved PL on the sharp transition at 0 V on the actuator and upon application of a voltage of 500 V. A fine-structure-splitting increase is observed as the voltage on the actuator is increased.

### 3.3.- Straining rare-earth and metal-ion doped thin films.

In metal-ion doped oxide films, the luminescence is due to the radiative decay of photoexcited electrons from high to low energy levels within the metal ions. The crystal field strength, which eventually depends on the strain fields present in the host material, can influence the electronic energy levels separation in the ions, therefore modifying the PL emission spectra. The systematic investigation of strain effects involves the fabrication of several samples in which a fixed crystal structure and ion doping is eventually obtained via chemical means. Unavoidable fluctuations in the fabrication process compromise the reproducibility of the samples and therefore the understanding of the mechanism underling the observed effect. In this regard, piezoelectric actuators offer an ideal solution to perform fundamental studies on a single sample.

In particular, Ni$^{2+}$-doped phosphors present a broadband near-infrared (NIR) emission from about 1200 to 1600 nm which makes it very promising for many applications like NIR optical amplifiers [116]. NIR luminescence tuning in SrTiO$_3$:Ni$^{2+}$ (STO:Ni) 600 nm-thick layers epitaxially grown on monolithic PMN-PT actuators has been demonstrated by introducing a reversible strain field (sketch of the device is shown in Figure 15a) [62]. The top electrode is realized by depositing an optically transparent indium tin oxide (ITO) layer. The NIR optical emission is assigned to the spin-allowed electronic transition Ni$^{2+}$:$^3$T$_2$($^3$P)→$^3$A$_2$($^3$F). The controlled introduction of a compressive deformation field in the material shortens the Ni$^{2+}$-O$^{2-}$ bond distance, and the enhanced crystal field induces a higher energy splitting between $^3$T$_2$ and $^3$A$_2$ levels. Figure 15b shows that when the voltage on the piezoelectric actuator is increased from 0 V to 500 V - a range corresponding to a compressive strain of 0.012% - the PL peak is blue shifted by about 13 nm, accompanied by a slight increase of the PL intensity. Both effects are explained by the enhanced crystal field around the Ni$^{2+}$ ions. Reversible PL shifts are observed by applying negative/positive voltages on the piezoelectric substrate, i.e. tensile/compressive regime (Figure 15c). Interestingly, independent of the sign of the strain field, a similar blueshift is reported due to a similar strain field magnitude. More details can be found in Ref. [62].

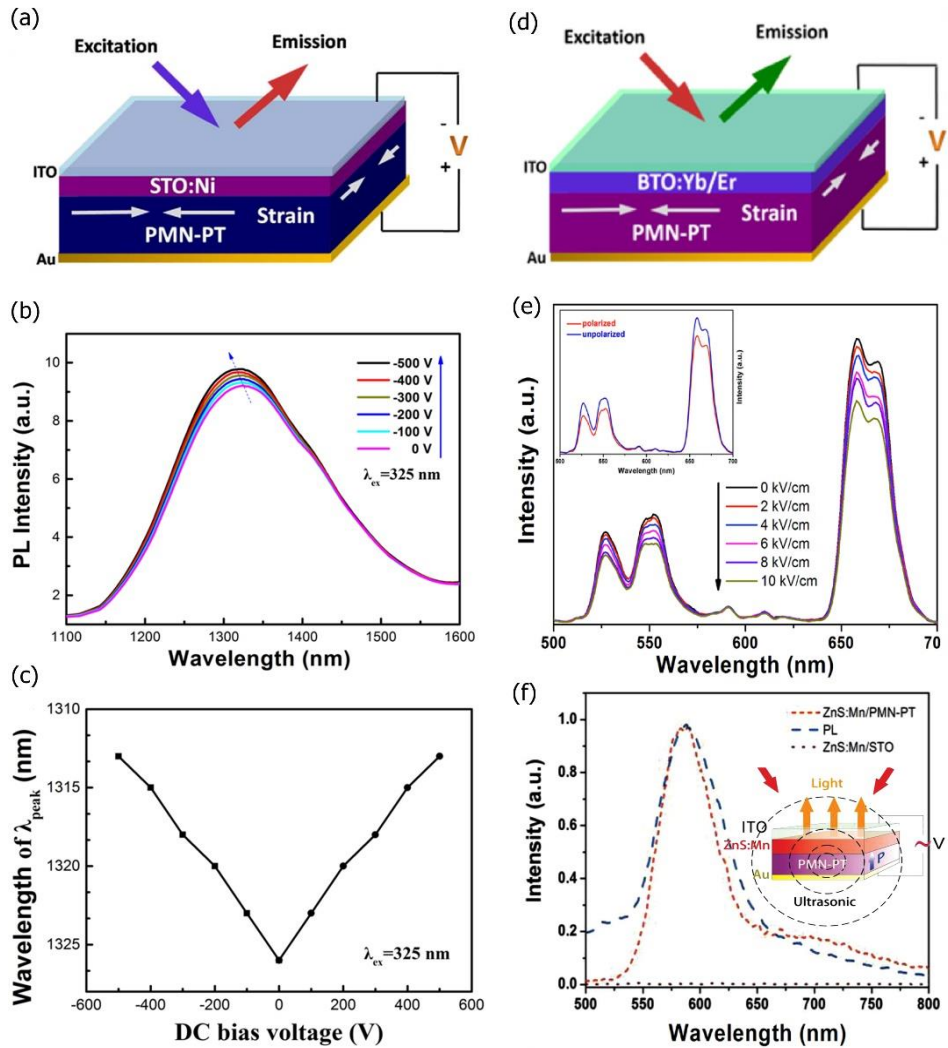

**Figure 15.-** (a) Sketch of a 600-nm-thick STO:Ni film epitaxially grown on top of a monolithic PMN-PT. An optically transparent ITO layer is grown on top of the structure as electrical contact. Reprinted by permission from Macmillan Publishers Ltd: [62] (doi:10.1038/srep05724). (b) PL spectra of the STO:Ni film for different applied voltages on the actuator. Reprinted by permission from Macmillan Publishers Ltd: [62] (doi:10.1038/srep05724). (c) Wavelength shift of the PL peak as a function of the applied voltage to the actuator. The tuning of the PL emission can be reversed in the tensile and compressive regimes. Reprinted by permission from Macmillan Publishers Ltd: [62] (doi:10.1038/srep05724). (d) Sketch of a BTO:Yb/Er film epitaxially grown on top of a monolithic PMN-PT. An optically transparent ITO layer is grown on top of the structure as electrical contact. Reprinted with permission from OSA from [66]. (e) PL spectra of the BTO:Yb/Er film for different applied voltages to the actuator in (d). Reprinted with permission from OSA from [66]. (f) Comparison of the luminescence in a ZnS:Mn film when: a voltage is applied to the actuator (red dashed line); PL signal when no voltage is applied to the actuator (blue dashed line). A sketch of the device is shown in the inset. Reprinted with permission from [67].

The effect of PMN-PT-induced biaxial strain on the upconversion PL processes in Yb/Er rare-earth ions doped BaTiO$_3$ (BTO) has been recently studied [66]. The device is fabricated following a similar one by growing epitaxially the BTO film on top of the PMN-PT substrate (Figure 15d), similarly to what discussed above. First, an ITO layer is grown as top electrode. The Yb$^{3+}$ ions are used as sensitizers for

Er$^{3+}$ ions in order to induce enhanced upconversion efficiencies. Here, the crystal symmetry of the host BTO material plays a crucial role in the PL observed from the dopants (rare-earth ions). In this case, the Er$^{3+}$ ions replace the Ti$^{4+}$ ions in the BTO crystal structure leading to a [ErO]$_6$ octahedron structure where Er$^{3+}$ ions are localized at the center. Due to lattice mismatch between the BTO film and the PMN-PT substrate, a tensile in-plane pre-strain is introduced in the film. The crystal symmetry around the Er$^{3+}$ is therefore lowered on the as-grown films, resulting in a higher PL intensity. By exerting stress (strain) fields with the piezoelectric actuator, it is possible to compensate the pre-strain obtained in the BTO film upon growth, leading to a PL intensity decrease, as can be seen in Figure 15e. Furthermore, it is possible to tune reversibly and dynamically the upconversion PL in the doped BTO films by applying dc/ac voltages with different polarities to the PMN-PT. As a proof of concept, the authors applied an ac voltage to the PMN-PT substrate and observed dynamical upconversion tuning for low frequencies.

The effects of strain on films presenting piezoelectric properties have been recently studied upon integration on piezoelectric actuators. In Ref. [67], ZnS:Mn thin films have been grown epitaxially on PMN-PT substrates by pulsed laser deposition. The PL of the ZnS:Mn layer presents a prominent band at around 588 nm resulting from the $^4T_1 \rightarrow {}^6A_1$ transition of Mn$^{2+}$ ions. ZnS doped with Mn$^{2+}$ (ZnS:Mn) possess piezoelectric characteristics and its band structure can be significantly changed by the piezoelectric potential induced upon application of a mechanical deformation in the material. In the absence of any optical excitation, the application of a voltage signal to the PMN-PT substrate leads to an optical emission, which is ascribed to the induced piezoelectric potential within the material as shown in Figure 15f. A similar PL signal is obtained when the film is optically excited. These observations can be explained considering that the inner-crystal potential can create an electron-hole pair with an electron from the valence band excited to the conduction band. The Mn$^{2+}$ ions can be promoted to an excited state following different mechanisms as described in Ref. [67]. The excited Mn$^{2+}$ ions can therefore return to the ground state by emitting light, where the induced luminescence is controlled by the state of deformation (strain) of the material, which can be in turn triggered by an electrical signal on the piezoelectric substrate. In particular, tuning of the emission intensity is possible through the application of a high-frequency ac voltage on the PMN-PT substrate up to the MHz regime. A higher enhanced PL intensity is observed for driving frequencies that match the mechanical resonances of the PMN-PT, in particular for the first mode at around 650 KHz.

**4.- Tunable optical properties of QDs by electro-elastic fields.**

In this section, recent works on the tailoring of the optical properties of QDs through the combined application of electric and strain fields are reviewed. The electric filed applied across a diode structure containing QDs can be used either to inject carriers in a LED structure (operated in the direct regime) or to tune the emission energy of the QDs (operated in the inverse regime). The latter opens many possibilities to tune the optical properties of QDs when combined with strain fields, as described below.

**4.1.- Stretchable Quantum-Light-Emitting diode.**

As mentioned above, µeV-control over the energy of the QD photons is one of the key requirements for most of the applications in quantum optics experiments involving more than one QD. One of the significant advances towards the implementation of QD-based devices with tunable optical properties are quantum-light-emitting diodes (QLED) integrated onto PMN-PT [18]. These devices consist of a p-i-n diode nanomembrane containing InAs QDs in the intrinsic region and bonded onto a PMN-PT monolithic actuator. The membranes are prepared and transferred by selective wet chemical underetching followed by gold thermo-compression bonding, as discussed in section 2.2.2. The device is operated by applying two voltages as *two independent tuning knobs*: one "knob" controls the current of the electric field ($F_d$) across the nanomembrane. The second "knob" (electric field, $F_p$ across the PMN-PT actuator), allows tailoring the strain field in the nanomembrane, as sketched in Figure 16a. In this device, gold is not only used as strain transfer-layer, but also as common ground for the two power supplies delivering $V_p$ and $V_d$, and as a back mirror of metal-semiconductor-dielectric ($SiO_2/TiO_2$) cavity used to enhance the light extraction efficiency. Remarkably, broad-band and tunable emission can be achieved without degrading the electrical operation of standard light emitting diodes [18]. Figure 16b shows a typical EL spectrum of the QLED, where X, XX and $X^-$ excitonic transitions from a single QD can be well resolved. All of them can be tuned by more than 20 meV by introducing a compressive/tensile strain field (up to 0.4%) when the electric field across the PMN-PT is swept in the range -15 to 55 kV/cm. The capability of this device to generate single photons in both tensile and compressive regimes is demonstrated by measuring the second-order correlation function on the X transition, leading to $g^{(2)}(0)<0.5$ in all the cases. Another interesting feature of this kind of devices is that they can be driven at relatively high frequencies by applying sub-nanosecond electrical pulses [117]. In this case of triggered single photon emission, the pulse signal (V~-0.7 V) is superimposed to a d.c. voltage of -1.7 V, which is just above the turn-on bias of the diode. Single photon emission is demonstrated under tensile and compressive strain at different frequencies from 80 MHz to 800 MHz with $g^{(2)}(0)$~0.1 (Figure 16c), thus proving the capability of the QLED to produce electrically-triggered and energy-tunable single photons.

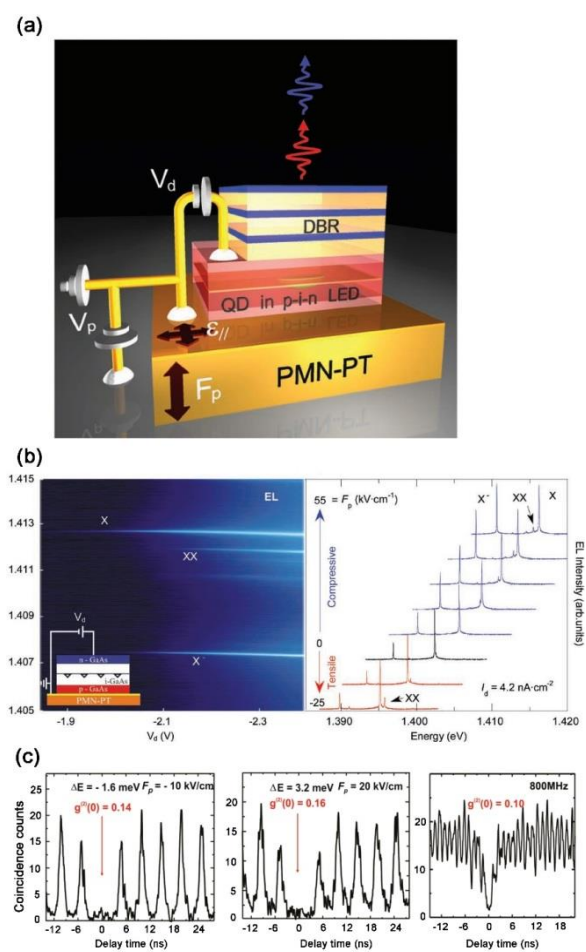

**Figure 16.-** (a) Sketch of a QLED device consisting of a p-i-n diode structure (with QDs in the intrinsic region) bonded on a monolithic PMN-PT actuator. A DBR structure consisting of alternated $SiO_2/TiO_2$ layers is deposited on top to complete a metal-semiconductor-dielectric cavity for boosting the flux of QD photons. The Au bonding layer is also used as electrical ground. Two independent voltages are applied for carrier injection into the QDs ($V_d$) and to drive the actuator ($V_p$) in order to strain the nanomembrane. Reprinted with permission from [18]. (b) Electroluminescence (EL) spectrum showing X, XX and $X^-$ excitonic transitions of a single QD. A significant energy shift in the compressive and tensile regimes is observed as the electric field $F_p$ on the actuator in varied. Reprinted with permission from [18]. (c) Autocorrelation $g^2(0)$ measurements demonstrating single photon emission from a single electrically-driven QD at 200 MHz (left and center panels) and 800 MHz (right panel). Adapted with permission from [117]. Copyright (2013) American Chemical Society.

**4.2.- Independent control of X and XX emission energies.**

Alternative proposals to produce entangled photon pairs from QDs – like the time reordering scheme [118] - requires a fine independent control over the emission energy of the biexciton and exciton complexes, and in particular on their energetic distance, i.e., the biexciton binding energy $E_b(XX)$. In QDs, the magnitude and sign of $E_b(XX)$ depends on the interplay between quantum confinement, direct Coulomb interaction, exchange and correlation effects and, in turn, on the QD structural details. While single external perturbations such as strain [119] and electric fields [79] can be used to tailor $E_b(XX)$ to some extent, the device discussed in the previous section can be used to achieve independent control of X and XX energies [55,119]. This is possible as strain and electric fields have a different effect on the interaction energies between carriers confined in a QD and affect $E_b(XX)$ in a different manner, as shown in Figure 17a. It is indeed apparent that $E_b(XX)$ shows a linear relationship with the X emission energy but with a different slope for strain and electric fields. Independent control of the emission energy and $E_b$ can be done by implementing a closed-loop system in which part of the collected light is used to provide feedback to the PMN-PT voltage supply [18]. First, the X emission energy is frequency stabilized to a target value ($E_{targ}$) by using the feedback loop on $F_p$ applied to the PMN-PT actuator, as shown in Figure 17b (bottom panel). Second, an electric field $F_d$ is applied to the diode to shift the XX via the quantum-confined-Stark-effect while the feedback loop on $F_p$ locks the X energy to $E_{targ}$. As strain and electric fields have a different effect on the X and XX complexes, the combination of the fields allows the binding energy to be modified for a fixed energy of the X transition.

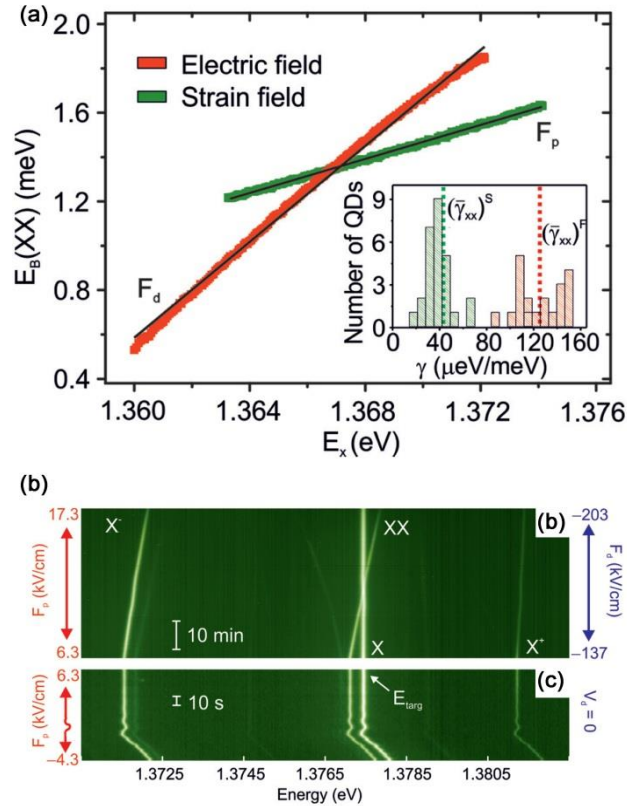

**Figure 17.-** (a) Biexciton (XX) binding energy against exciton (X) emission energy measured on a nanomembrane containing QDs. A linear relationship is found for both an electric field applied on the actuator and on the diode structure (reverse bias). The slopes of the fitted experimental data are shown in the inset. (b) The exciton X energy is locked to a target value by driving the electric field on the actuator (bottom panel). The biexciton XX energy is tuned into coincidence with the X by applying simultaneously an electric field on the diode structure and piezoelectric actuator (up panel). The latter is corrected accordingly in order to keep a constant target X energy emission. Reprinted (figure 1) from [55]. Copyright (2013) by the American Physical Society.

**4.3.- Piezotronic effect in QD membranes.**

In a recent publication, swapping the position of the electron and hole wave functions in a QD has been experimentally reported by introducing anisotropic strain fields [120]. The device employed in these experiments consists of a p-i-n diode nanomembrane containing InAs QDs bonded on a monolithic piezoelectric device by gold thermocompression bonding, as sketched in Figure 18a. The QDs are embedded in a GaAs/AlGaAs quantum well to suppress tunneling at high electric fields and to enable broad-band tuning of the QD emission lined via the quantum confined Stark effect. Figure 18b shows µ-PL spectra from a single QD as a function of the electric field across the diode structure in the inverse regime. The energy shift of the transitions induced by the applied electric field is governed by the Equation (1)

$$E = E_0 - pF_d + \beta F_d^2 \qquad (1)$$

where $E_0$ is the transition energy at $F_d = 0$ and $p = ez$ the built-in dipole moment ($e$ is the elementary charge and $z$ is the electron-hole separation along the crystal growth direction). The parameter $\beta$ is the polarizability. The electric field is given by $F_d = (V_d + V_{bi})/d_d$, where $V_d$ is the applied voltage, $d_d \sim 150\ nm$ is the thickness of the nanomembrane and $V_{bi}$ is the built-in voltage. The latter can be estimated from the I-V trace of the diode, as shown in Figure 18c. It turns out that there is a gradual shift of built-in voltage against $F_p$, i.e., against the stress applied by the actuator. The magnitude of this shift is roughly a factor 10 larger than the one expected by the strain-induced changes of the band gap energy, and can only be understood considering piezoelectric effect. More specifically, piezoelectric charges that are generated when $F_p$ is applied lead to a modification of Schottky barrier present at the interface between the p-i-n diode and the Au contact on the nanomembrane (see figure 18) that, in turn, modify the onset of the I-V trace. Taking advantage of this so-called "piezotronic" effect [121,122], it is possible to estimate the in-plane stress delivered by the actuator, which is found to be highly anisotropic, with $|S_1|/|S_2|\sim 3.6$ and $S_1 + S_2 \sim -180$ MPa, with $S_1$ and $S_2$ the principal major and minor stresses, respectively.

Equation (1) can be used to fit the experimental data reported on Figure 18b and to extract the dipole moment ($p$) vs the applied stress (see Figure 1d). Remarkably, as the stress is varied the electron-hole dipole moment changes not only in magnitude but also in sign, meaning that strain can be used to swap the position of the electron and hole wave functions along the crystal growth direction (see Figure 18d). By performing detailed theoretical calculations based on **k·p** models and by using the estimated stress configuration, it is concluded that the inversion of the exciton dipole moment is driven by the piezoelectric fields. Moreover, the calculations reveal that nonlinear terms in the piezoelectric fields have to be taken into account to correctly reproduce the experimental findings. Therefore, the inversion of the exciton dipole moment is one of the first experimental evidence of the importance of nonlinear piezoelectric effects in III-V semiconductors [123–125].

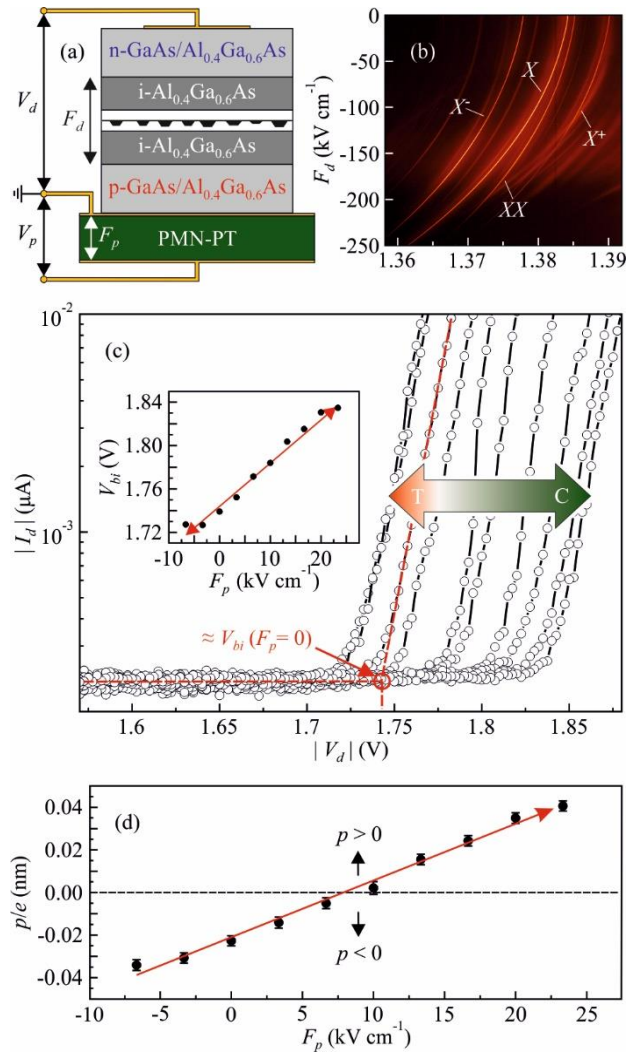

**Figure 18.-** (a) Sketch on a p-i-n diode nanomembrane containing QDs bonded on a PMN-PT monolithic actuator. Out-of-plane electric fields and in-plane strain fields can be introduced by applying a voltage ($V_d$) to the diode or to the actuator ($V_p$), respectively. (b) Energy shift of exciton and multi-exciton complexes as the electric field on the diode is varied. (c) Diode I-V trace shift as the strain field introduced by the actuator is varied in the tensile and compressive regimes. The inset shows the extracted diode built-in potential ($V_{bi}$) for each value of the electric field applied on the actuator ($F_p$). (d) Dipole moment of excitons confined in a QD as a function of the electric field $F_p$. The magnitude of the dipole can be changed, as well as its sign, by varying the electric field on the actuator. Adapted from Ref. [120].

## 4.4.- Erasing the FSS using diode-like membranes integrated onto monolithic piezoelectric actuators.

Since the in-plane piezoelectric constants of ideal PMN-PT (001) substrates are equal, a nearly isotropic biaxial strain field is expected to be transferred to the bonded semiconductor nanomembranes. However, anisotropic biaxial strain fields are systematically observed experimentally [18,33], especially on devices processed by gold thermo-compression bonding. There are several factors that lead to a deviation from the expected behavior. On the one hand, commercially available PMN-PT substrates are polycrystalline with slight misorientation between different crystallites. On the other hand, other aspects related with the bonding interlayer such as quality of the bonding (as pointed out in section 2.2) can introduce unexpected and unpredictable non-uniformities in the bonding layer like air gaps, which affect both magnitude and anisotropy of the applied in-plane stress field in the membrane. This unexpected behavior has important consequences in the tailoring of the inherent properties of QDs, as discussed in the previous section and below.

In Ref. [33] it has been shown that the coherent coupling of the bright states in QDs (and in turn the FSS or *s*) can be reversibly controlled and erased by combining vertical electric fields along the [001] direction ($F_d$) and an in-plane anisotropic biaxial stress field with anisotropy $\Delta S = S_1 - S_2$ ($F_p$ on PMN-PT), being $S_1$ and $S_2$ the magnitudes of two major and minor principal stresses, respectively, applied along arbitrary directions in the (001) plane [33].

Anisotropic strains (with principal directions different from the [110] and [1-10] crystal directions of GaAs) and electric fields are always sufficient to remove the FSS because they allow two different QD parameters to be tuned independently: the polarization direction of the exciton emission (ϕ) and the magnitude of the fine structure splitting (*s*) (both can be easily measured performing standard polarization-resolved micro-PL measurement). For the case of an electric field ($F_p$) applied along [001] direction and biaxial stress fields, the effective two-level Hamiltonian for the bright excitons is given by Equation (2) [19,82]:

$$H = [\eta + \alpha \Delta S + \beta F_d]\sigma_z + [k + \gamma \Delta S]\sigma_x \qquad (2)$$

where $\sigma_{x,z}$ are the Pauli matrices and $\eta$ and $k$ account for the QD structural asymmetry. The effect of the external fields are considered in the Hamiltonian through the parameters $\alpha$ and $\gamma$ (related to the elastic compliance constants renormalized by the valence band deformation potentials) and $\beta$ (proportional to the difference of the exciton dipole moments). Upon diagonalization of Equation (2), the following values are obtained for *s* and ϕ:

$$s = [(\eta + \alpha \Delta S + \beta F_d)^2 + (k + \gamma \Delta S)^2]^{1/2} \qquad (3)$$

$$tan\phi_\pm = \frac{k+\gamma p}{\eta+\alpha\Delta s+\beta F_d \pm s} \qquad (4)$$

It can be seen that Equation (3) is minimized for specific $F_d$ and $\Delta S$ values fulfilling the following equations:

$$\Delta S_{critic} = -\frac{k}{\gamma} \qquad (5)$$

$$F_{d,critic} = \frac{\alpha k}{\gamma\beta} - \frac{\eta}{\beta} \qquad (6)$$

It is therefore possible to find specific values for the in-plane stress ($F_p$) and out-of-plane electric fields ($F_d$) such that s=0 is accomplished. This is demonstrated experimentally in Figure 19, which shows the experimental data obtained for *s* and ϕ for different $F_p$ values while varying the electric field $F_d$. The solid lines are fits to the experimental data using Equation (3) and (4). An excellent agreement is obtained, and also the change in handedness of ϕ as expected upon crossing of the s=0 critical point is reproduced. Remarkably, the proposed device allows to reshape the electronic structure of any arbitrary QD so that a value *s*=0 can be always reached. In turn, this suggests that any arbitrary QD can be used to generate highly polarization entangled photons, an experiment which was performed in Ref. [75]. It is important to emphasize that, from a theoretically point of view, the two perturbations (strain and electric fields) are needed to control – via ϕ and *s*, see Equation (3) and (4) – the two QD parameters $\eta$ and *k* that are responsible for the non-zero value of the FSS in as-grown QDs [78]. From an experimental point of view, on the other hand, the condition s=0 can be realized by using the electric field $F_d$ to align the polarization axis of the X emission (ϕ) along the actuating direction of the anisotropic stress field [82]. This also implies that the anisotropic strain field alone is sufficient to suppress the FSS, provided that ϕ of the as-grown QDs is already aligned along the major or minor stress axis. However, each QD has slight different ϕ and the electric field (or another tuning knob) is needed to reach the s=0 condition. This discussion not only explains why anisotropic strain fields are so effective in tuning the FSS, but it also suggests a different way to reach the s=0, as discussed below.

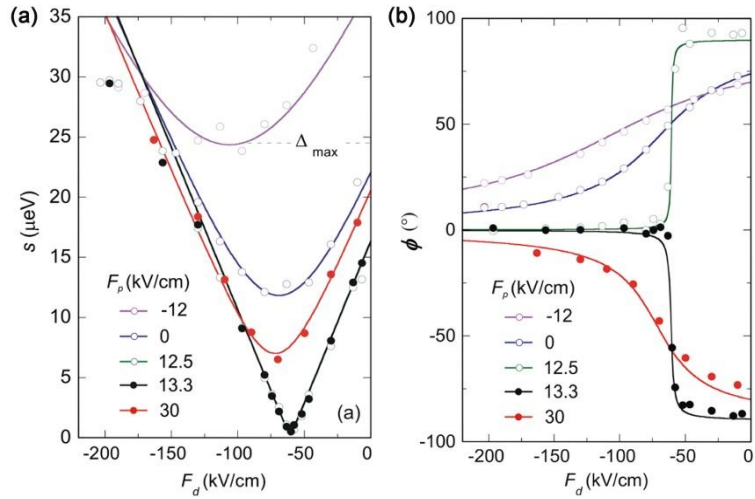

**Figure 19.-** (a) Fine-structure-splitting values (*s*) and (b) polarization angle (φ) for the exciton X of a QD as a function of the electric field applied on a p-i-n diode containing QDs ($F_d$) for different electric fields applied on the PMN-PT actuator ($F_p$). At certain critical values $F_{d,critic}$ and $F_{p,critic}$ (anisotropic biaxial stress field $\Delta S_{critic}$), it is possible to tune *s* to zero. The experimental data is represented with empty(full) circles for $F_p$ values smaller(larger) than 13 kV/cm. The solid lines represent the fits to the experimental data using Equation (3) and (4) in the main text. Reprinted (figure 3 and figure 4) from [33]. Copyright (2013) by the American Physical Society.

The in-plane anisotropies in the confining potential that are responsible of the value of φ arise from the QD structural asymmetries, which are QD specific. However, it is known that self-assembled InAs QDs on GaAs(001) substrates are often preferentially elongated in shape along the [1-10] GaAs crystal direction. As a consequence, an *average* polarization angle for the X close to this direction is expected. This peculiarity was used in Ref. [126] to effectively tune the FSS to very small values by simply aligning the [110] direction of the GaAs nanomembrane containing the QDs to the actuating direction of the piezoelectric crystal. PL polarization resolved measurements on several InAs QDs on PMN-PTs show preferential polarization angles of the X close to [1-10] when no strain field is applied, corresponding to φ~90° in Figure 20a. This result is obtained on approximately 30% of the measured QDs. It should be noted that the piezo used in this work is a PMN-PT (011) oriented substrate, which provides highly anisotropic strain fields with in-plane anisotropy $\varepsilon_{xx} \approx -0.37\varepsilon_{yy}$ due to very different piezoelectric coefficients $d_{31}$=+420 pCN$^{-1}$ and $d_{32}$=-1140 pCN$^{-1}$ along [0-11] (x axis) and [100] (y axis) directions of the crystal, respectively. Hence, a GaAs nanomembrane was bonded on a PMN-PT (011) substrate by carefully aligning the GaAs crystal axes [1-10] and [110] along the [100] and [0-11] of the PMN-PT (011) substrate, respectively. Figure 20b shows the *s* values for different QDs as the electric field on the actuator is swept. It is apparent that the minimum value of the FSS is achieved for the QD that is closely aligned along the [1-10] and that even slight misalignments of the X polarization angle induce significant changes in the minimum value of $s$ that can be reached when sweeping $F_p$. The X polarization angle is rotated in a similar way as discussed above (Figure 20c). Since the minimization of $s$ is achieved using only strain, the electric field across the diode can be employed to electrically inject carriers in the QDs at different frequencies so as to trigger the generation of entangled photons in an entangled-light-emitting-diode (ELED) device. Tomography measurements on XX and X photons from QDs featuring FSS=0.6 µeV upon the adequate application of a strain field show a concurrence value of 0.69±0.04. The generation of such entangled photons is demonstrated under pulsed electrical injection with frequencies up to 400 MHz.

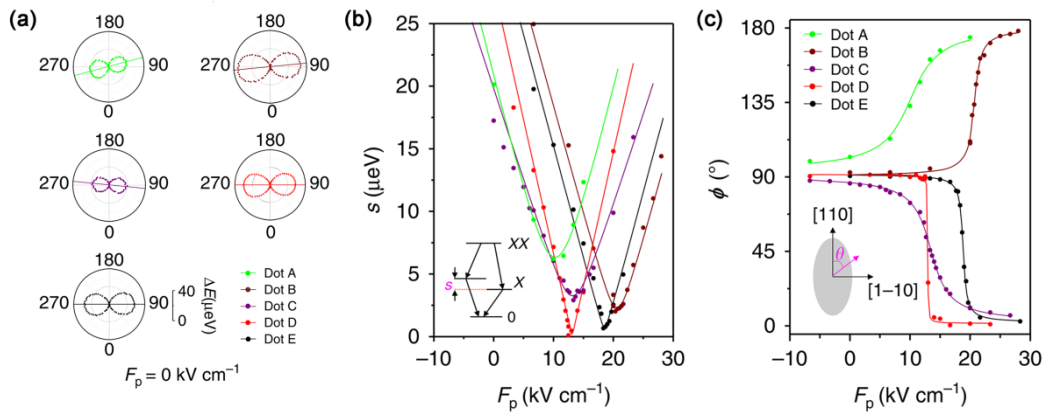

**Figure 20.-** (a) Polar plot of polarization resolved PL measurements on several QDs. A clear alignment along the [1-10] crystal direction of the GaAs matrix is found in about 30% of the investigated QDs. (b) Fine-structure-splitting ($s$) and (c) X polarization angle ($\phi$) values for the same QDs as a function of the electric field on the PMN-PT actuator. Reprinted from [126].

Although this work represents a significant step towards the development of electrically-driven sources of entangled photons from QDs, it requires finding QDs with the right exciton polarization angle. Unfortunately, this depends on many experimental factors like growth protocols and/or bonding procedures to transfer the membrane on the actuator. In addition, generation of entangled photons with tunable emission energies are mandatory for advanced quantum experiments, a task that cannot be accomplished easily with the dual-knob device discussed so far. At this point, it is clear that an additional tuning knob is needed in order to deterministically tune the energy of entangled photons emitted from

QDs. To tackle this issue, a novel class of micro-machined piezoelectric actuators capable of exerting in-plane strain fields with controllable magnitude and anisotropy is presented in section 5.1.2. of this manuscript.

**5.- Novel micro-machined piezoelectric actuators.**

Monolithic actuators are suitable for a broad range of applications where nearly bi-axial strain fields with magnitudes ~ 0.1% are required, as discussed in detail in the previous section. In order to fully exploit the full potential of elastic strain engineering of nanomaterials, it is highly desirable to attain a full control of the magnitude, direction and anisotropy of the exerted strain fields. Advanced techniques such as femtosecond laser cutting allows the micro-machining of the piezoelectric substrates with arbitrary designs. In this section, we will introduce a new class of micro-machined piezoelectric actuators that allow the in-plane strain tensor in semiconductor nanomembranes to be fully controlled. The micro-machined actuator essentially consists of a piezoelectric substrate cut by the femtosecond laser technique. Depending on the specific application, it can feature a different number of actuating micron-sized legs. The simplest design consists of two micro-sized legs disposed in front of each other that, upon bonding of a nanomembrane on top of the legs, can introduce nearly uniaxial stress fields with relatively high magnitudes, as discussed in section 5.1.1. For certain applications where full control of the in-plane stress tensor in the bonded nanomembrane is required, a 6-legged actuator can be used as discussed in section 5.1.2. The capabilities of the latter for the generation of polarization-entangled photons with tunable energy from self-assembled InAs QDs is demonstrated (section 5.2). Advantages and disadvantages of the approaches are discussed in view of the targeted applications.

**5.1.- Micro-machined piezoelectric actuators for full in-plane stress fields control in GaAs nanomembranes.**

**5.1.1.- 2-legged device.**

Figure 21a shows a sketch of GaAs nanomembrane bonded a 2-legged actuator coated with Au (side A, electrically grounded). The micro-machining of the PMN-PT piezoelectric actuator is performed using a femtosecond laser with 350 fs pulse duration, 25 KHz repetition rate and 6 µJ pulse energy. The sample employed in this work consists of a [GaAs(001) substrate/100-nm-thick $Al_{0.7}Ga_{0.3}As$ sacrificial layer/330-nm-thick GaAs] layered structure grown by MBE. The fabrication procedure is similar to the case of a monolithic device: the sample is bonded on the actuator (gold thermo-compression bonding) by carefully aligning the [110] GaAs crystal direction with respect the [100] direction of the PMN-PT actuator (direction defined by the two legs) followed by a selective wet chemical etching to remove the whole substrate. On the other side (side B), only the legs are coated with Au so as to apply independent voltages on each of the legs. The photoluminescence (PL) measurements are performed in the suspended GaAs region between the legs on an area of around 10 µm (size of the laser spot). The working principle of the device is based on the application of a voltage on the two micro-machined legs, while the top side is set to ground (i.e. an electric field is induced across the piezoelectric material). This produces an out-plane deformation on the piezoelectric legs – in-plane deformation due to the Poisson effect – that in turn introduces an in-plane strain field on the bonded nanomembrane. The tensile/compressive character of the strain field can be controlled by the sign of the applied voltage and by the poling direction of the piezoelectric substrate.

Semiconductor GaAs is a well-known direct band-gap material with bright PL and known response to deformations. More specifically, the polarization properties of light emitted by the recombination of light- and heavy-hole free excitons in GaAs can be used as a stress gauge. Therefore, standard polarization-resolved micro-PL measurements combined with k.p theory can be used to estimate the strain status of a GaAs nanomembrane subject to arbitrary stress fields, with a spatial resolution of the of order of 1 µm. Since the piezoelectric actuator exerts stress only in the (001) plane of the GaAs crystal, only the in-plane stress is relevant here. The stress status can therefore be fully characterized by the three components of the stress tensor ($S_{xx}$, $S_{yy}$, $S_{xy}$) or, equivalently, by the principal major ($S_1$)/minor ($S_2$) stresses and an angle ($\theta_{S1}$) between $S_1$ and an arbitrary reference direction (GaAs [110] crystallographic direction in this case). This direction is also taken as a reference for the PL polarization-resolved measurements.

Unstrained GaAs presents degenerate light-hole (LH) and heavy-hole (HH) valence bands at the Γ point of the Brillouin zone and, upon the recombination of light- and heavy-hole free excitons, a single unpolarized PL emission peak should be ideally observed. On the other hand, an in-plane stress (strain) breaks the crystal symmetry and removes the degeneracy between the two, leading to two emission peaks. Polarization-resolved PL maps for different voltages applied on both legs are shown in Figure 21b. The PL spectra show two PL peaks indicating that the nanomembrane at the point of measurement (gap between the legs) is pre-strained, which is likely due to bonding and device processing. Since in the presence of arbitrary in-plane strains the two observed transitions involve an admixture of HH and LH bands, the low and high energy PL peaks are referred as $E_1$ and $E_2$, respectively. The splitting in energy between the two free excitons is not the only experimental observable that allows the strain status of the membrane to be estimated. In fact, because of band mixing, the light collected along the perpendicular direction to the membrane is elliptically polarized for both $E_1$ and $E_2$ (see Figure 22d), where the polarization angle of the $E_1/E_2$ components corresponds to the direction of the minor/major stress axes. Hence, the stress state is fully encoded in the energies $E_1$, $E_2$ and the polarization angle ($\varphi$). First, the corresponding direct problem is solved by calculating the expected PL spectra for a given stress configuration. This is performed by making use of the 8-band k·p theory where the stress is introduced via the Pikus-Bir Hamiltonian, and the dipole approximation is used for the treatment of the optical properties. The Hamitonian is diagonalized at the point of the Brillouin zone where the minimum/maximum of the conduction/valence bands are located (Γ point), respectively, to obtain the single-particle energies of electrons and holes and the corresponding eigenvalues. The latter is used to extract the transition probabilities as a function of the polarization angle.

The experimentally measured energies $E_1$, $E_2$ and polarization angle $\varphi$ are then used as input data for a non-linear least-square minimization algorithm capable of retrieving the stress state $(S_1, S_2, \theta_{S1})$ [29] of the nanomembrane which minimizes the deviations between measured and calculated $E_1$, $E_2$, and $\varphi$ values. The first guess for the fit is obtained from a semi-analytical solution of the Pikus–Bir Hamiltonian without inclusion of the split-off band. The split-off band is then included during the least-square minimization. Self-consistency and uniqueness of the solution were thoroughly checked.

In the case of the 2-legged actuator presented in Figure 21, the calculated pre-stress value at zero voltage is characterized by principal stresses $S_1=150$ MPa and $S_2=76$ MPa with $S_1$ forming an angle $\theta_{S1}=62°$ with the [110] crystal direction. An anticrossing between the two free-excitons (with a minimum energy splitting of about 6 meV) is obtained, indicating that the induced stress is not aligned with the pre-stress axes and therefore, the unstrained configuration cannot be reached with these devices (Figure 21c). The induced major stress angle with respect to the [110] GaAs crystalline direction as a function of the applied voltage is shown in Figure 21d. It should be noted that the pre-strain present in the suspended nanomembrane at the gap between the legs upon processing of the device can induced a slight arbitrary bending of the membrane that should be first compensated. The latter is observed when the major stress angle approaches a zero value (uniaxial character) for voltages larger than 100 V (tensile stress fields), which is the regime where the bending of the nanomembrane is negligible. The angle values dispersion observed for negative voltages (compressive stress fields) is ascribed to an enhanced buckling/bending of the nanomembrane. In this regime, assuming bending of the nanomembrane, the induced stress cannot be properly controlled by the device. The calculated induced hydrostatic stress ($S_1+S_2$) and anisotropy ($S_1-S_2$) for each of the voltages are shown in Figure 21e. The geometry and gaps between the micro-machined actuating legs eventually condition both tunability of the stress field anisotropy and its magnitude, respectively. Specifically, the latter depends on the ratio between the length of the actuating legs and the gap between them. Hence, by keeping a constant legs length (~1500 µm), the achievable stress magnitude uniquely depends on the gap between the legs. In the case of the 2-legged actuator, a highly anisotropic stress up to $S_1-S_2=263$ MPa and hydrostatic stress up to $S_1+S_2=419$ MPa are calculated for the tensile regime (400 V). The deviation from a purely uniaxial stress is attributed to a possible non-uniform bonding on top of the legs. Remarkably, these values are larger than those obtained for the 6-legged device even for a much lower electric field on the PMN-PT actuator (13 kV/cm), as shown in section 5.1.2. This is due to a smaller gap between the legs in the case of the 2-legged actuator (~80 µm) with respect to the 6-legged actuator (~140 µm).

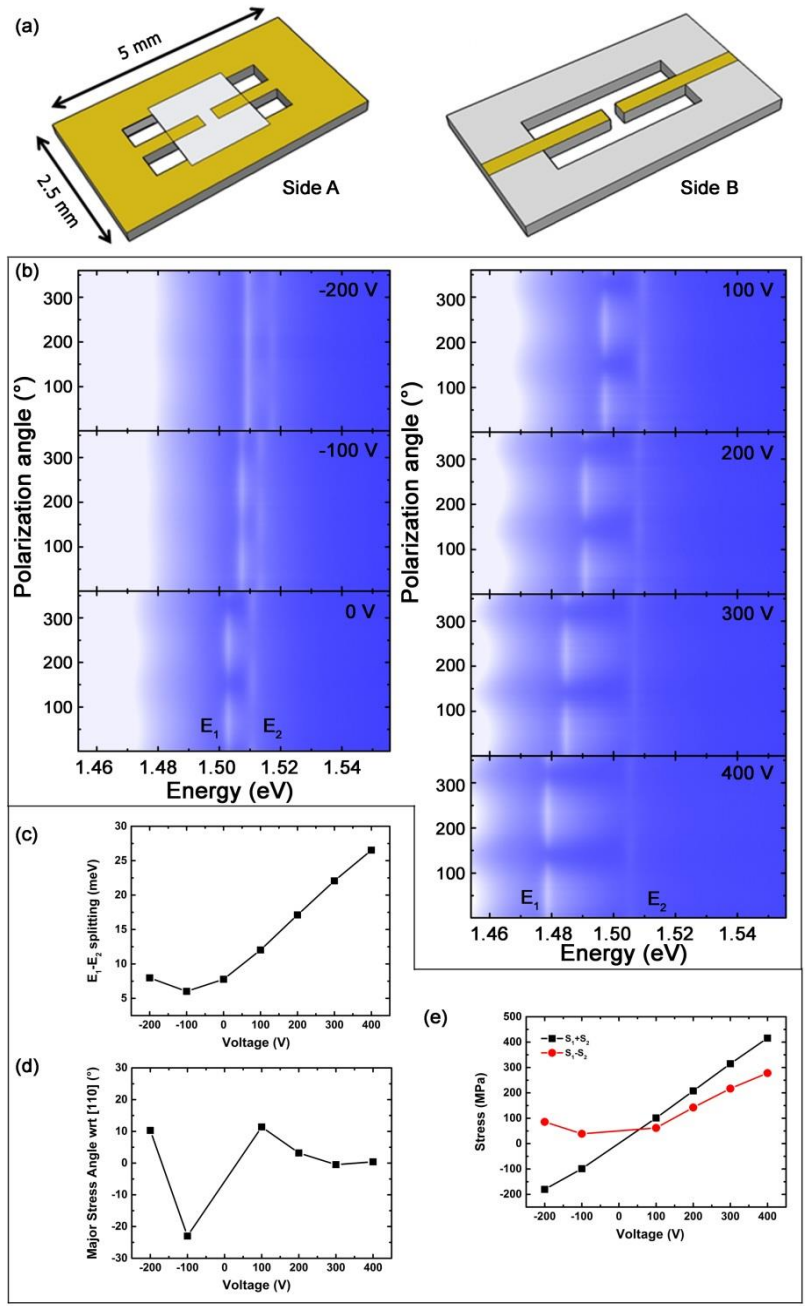

**Figure 21.-** (a) Sketches of the 2-legged device. A GaAs nanomembrane is bonded on the top side (side A) of the actuator. This side is coated with gold and set to ground. The contacts for applying the voltage on the two legs are defined at the bottom side (side B). (b) Color-code PL spectra as a function of the polarization angle for several voltages applied simultaneously on both legs. (c) Energy splitting $E_1$-$E_2$ as a function of the applied voltage. (d) Calculated induced major stress angle ($\theta_{S1}$) as a function of the applied voltage. (e) Calculated induced hydrostatic stress and anisotropy as a function of the applied voltage.

Figure 22 shows the PL emission shift of the exciton X and multi-exciton (MX) transitions from a GaAs QDs embedded in a AlGaAs nanomebranes integrated onto a 2-legged piezoelectric device. A total PL shift of 41.5 meV is found for this particular device and for a range of applied electric fields up to 22 kV/cm. Interestingly, due to the large optical emission tunability of such device, the optical emission of as-grown GaAs QDs can be tuned reversibly through both $D_1$ and $D_2$ transitions which is very interesting for applications in storage of single photons in thermal Rb atomic clouds [85,86]. (see also section 3.1.4.). Recently, a QDs based atomic quantum memory at the Rb $D_1$ line has been successfully demonstrated [127].

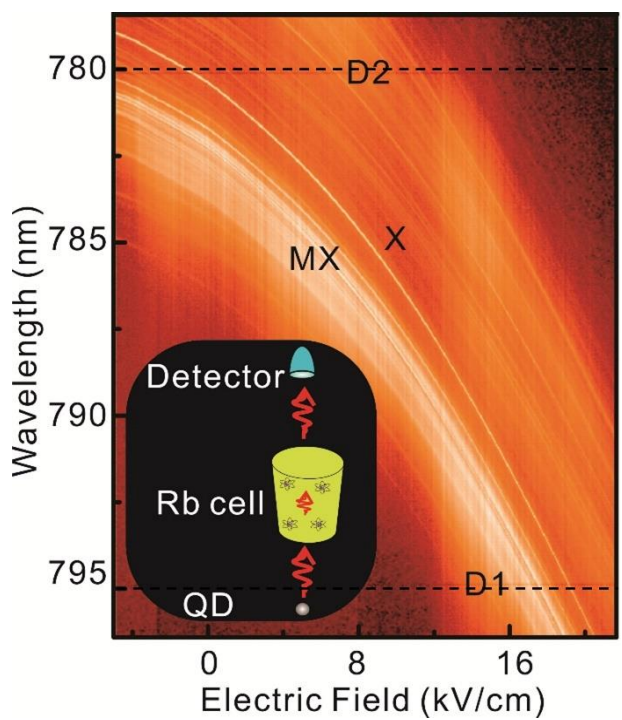

**Figure 22.-** PL emission wavelength from a GaAs QD as a function of applied voltage (electric field) on a 2-legged actuator (X and MX refer to exciton and multi-exciton transitions, respectively). The inset shows a potential application of the device for tuning the QD optical emission wavelength in resonance with a vapor of Rb atoms contained in a cell (yellow cylinder). The emission can be tuned through $D_1$ and $D_2$ Rb transitions.

**5.1.2.- 3-legged and 6-legged devices.**

In Ref. [29], a novel class of piezoelectric actuators that are capable of controlling the three components of the in-plane stress tensor in semiconductor nanomembranes is demonstrated. Most importantly, arbitrary stress configurations including uniaxial and isotropic/anisotropic biaxial stress fields along any direction can be produced. In the following, a variety of micro-machined actuators featuring different designs are presented and their suitability to deliver arbitrary stress fields with relatively high magnitudes is discussed. Generally speaking, these actuators allow exerting strain fields at least one order of magnitude higher than those obtained with the monolithic actuators presented in the previous sections.

The first example we discuss here is a "3-legged device", i.e., a micro-machined piezoelectric actuator featuring three legs oriented radially along directions 60° away from each other, and with a GaAs nanomembrane bonded on its top surface (see Figure 23a). The legs have a 300x300 μm² cross-section and a length of 1500 μm.

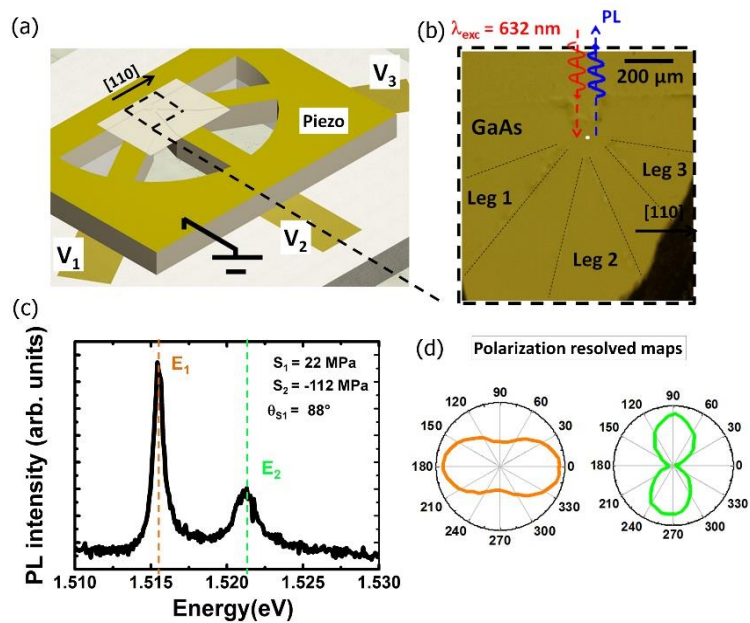

**Figure 23.-** (a) Sketch of a 3-legged device consisting of a micro-machined PMN-PT substrate with a GaAs nanomembrane bonded on its top surface. Three independent voltages ($V_1$, $V_2$, $V_3$) are applied to individual legs at the bottom of the actuator by contacting each of the legs on an AlN chip carrier through gold pads defined by optical lithography on the chip carrier. The top side of the actuator is gold coated and electrically grounded. (b) Optical picture of the GaAs nanomembrane bonded on top of the three piezoelectric legs. The PL signal is measured at the central gap between the legs, see white circle. (c) Low temperature (10 K) PL spectrum of the GaAs nanomembrane at zero applied voltage. The anisotropic pre-strain lifts the degeneracy of the valence bands leading to two PL emission peaks at energies $E_1$ and $E_2$. (d) PL intensity for peaks $E_1$ and $E_2$ as a function of the polarization angle. Adapted with permission from [29].

Achieving full control of the in-plane stress field means that the three stress field components can be controlled independently. This can be accomplished by employing three independent "tuning knobs" capable of generating three independent stress configuration whose axis are not all parallel to each other. In our device, these knobs are provided by the three micron-sized piezoelectric legs. In this case, the working principle of the device is based on the application of three independent voltages ($V_1$, $V_2$, $V_3$) on the three micro-machined legs, while the top side is set to ground.

Figure 23b shows an optical image of the bonded GaAs nanomembrane in which the underlying three piezoelectric legs can be distinguished. The PL signal is measured at the central gap between the three legs (white point corresponding to an area with a diameter of ~10 μm). A cw HeNe laser (632.8 nm) is used for excitation. The PL spectrum at 10K at the central gap is shown in Figure 23c for no voltages applied to the piezo-legs. This observation clearly indicates the presence of pre-strain in the nanomembrane, likely due to bonding and device processing (see section 5.1.1).

Assuming linearity in the elastic strain theory, the total stress state can be expressed as:

$$\bar{S}_t(\bar{V}) = \bar{S}_{pre-stress} + \bar{s}_{induced}(\bar{V}) = \bar{S}_{pre-stress} + \bar{\bar{T}} \cdot \bar{V} \qquad (7)$$

where $\bar{S}_t = (S_{xx}, S_{yy}, S_{xy})$ are the stress tensor components in Voigt notation, $\bar{S}_{pre-stress}$ the stress at zero applied voltage, and $\bar{\bar{T}}$ is a 3×3 "transfer matrix". The stress induced by the actuator can be then calculated by subtracting the pre-stress in the membrane to the total stress ($\bar{s}_{induced} = \bar{S} - \bar{S}_{pre-stress}$).

Figure 24a shows the corresponding PL spectra shift vs. $V_3$ while keeping the voltages applied to the other legs to zero. From the splitting of the two free exciton at $V_3=0$ it is possible to estimate a pre-stress of $S_1$=-8±1 MPa, $S_2$=-78±2 MPa, $\theta_{S1}$ =110±1°. When $V_3$ is varied, a clear anticrossing between the two free-excitons is observed (with a minimum splitting value of 2.5 meV), a clear signature that leg 3 is not able to compensate alone for the pre-stress arising during device fabrication. However, by properly tuning $V_1$ and $V_2$ it is possible to reach a configuration in which the stress field exerted by the third leg is applied along the same direction of the pre-stress axes. In this case, the symmetry of the crystal can be fully recovered and a crossing of the PL components is observed as $V_3$ is swept (Figure 24b). As expected, unpolarized light is measured at the crossing point. Moreover, a 90° rotation of the polarization angle is found on both PL components when the stress is tuned through the crossing point.

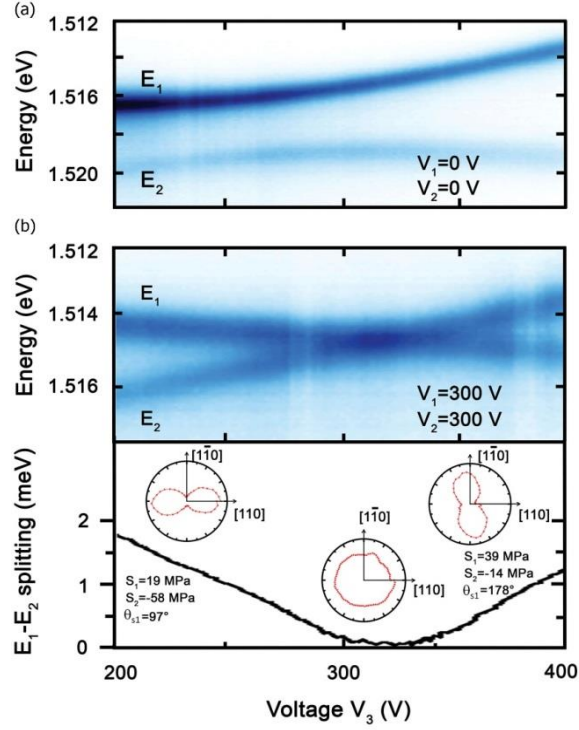

**Figure 24.-** (a) Color-code PL spectra of GaAs as a function of the voltage on leg-3 ($V_3$). The voltages on leg-1 ($V_1$) and leg-2 ($V_2$) are kept to zero. An anticrossing of the $E_1/E_2$ components is revealed. (b) Same as in (a) when $V_2$ and $V_3$ are set so that the actuating direction of leg 3 is aligned along the pre-stress axis. A crossing of the the $E_1/E_2$ PL components is observed, indicating recovery of the unstrained state in GaAs. The polarization angle for $E_1$ and $E_2$ PL components rotate by 90° when the stress field is tuned through the crossing point. Unpolarized PL emission is measured for unstrained GaAs at the crossing point. Adapted with permission from [29].

The capability of the device to compensate the non-controllable pre-stress arising during device processing is a very first indication of the possibility to exert deliberate in-plane stress fields on demand. In order to prove this statement, it is sufficient to use the linearity of the elastic theory, and in particular Equation (7). To do that, the device behavior has to be first calibrated to obtain the transfer matrix, which then allows us to calculate the corresponding voltages to be applied on each of the legs so as to achieve the desired strain configuration (encoded via $S_1$, $S_2$, $\theta_{S1}$). Practically, this is done by performing PL measurements while sweeping independently the voltages applied to each leg. The PL data in combination with Equation (7) are used to calculate the transfer matrix coefficients and then the induced stress field via $\bar{s}_{induced} = \bar{\bar{T}} \cdot \bar{V}$. The capabilities of the device to procedure stress fields on demand is demonstrated by predicting a number of induced stress fields with variable magnitudes and anisotropies on the GaAs nanomembrane. We refer the reader to Ref. [29] for detailed information on this procedure. Comparative data between experimental data and theory for a different polarization angle is shown in

Figure 25 by sweeping the voltages on each of the legs in order to induce a uniaxial stress up to 100 MPa (major induced stress s₁) along a fixed direction of $\psi_{s1} = 15°$ with respect to the [110] direction.

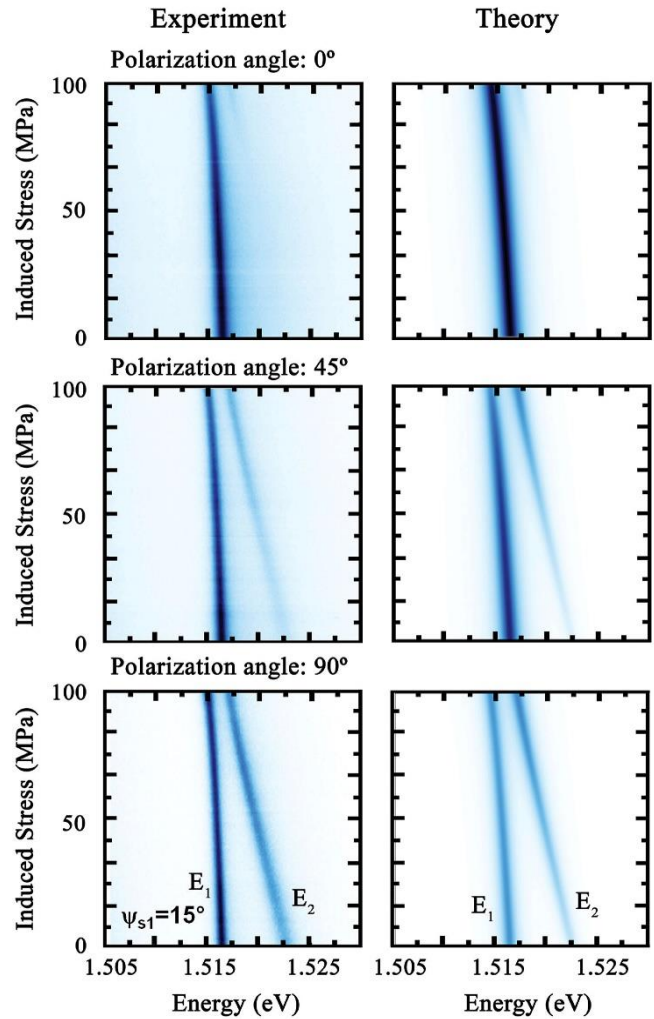

**Figure 25.-** Color-coded micro-PL spectra for three different polarization directions at 0°, 45° and 90°. The voltages on each of the legs are swept to induce a uniaxial induced stress field up to 100 MPa along a fixed principal major stress direction of 15° with respect to GaAs [110] direction. The acquired experimental data (left panels) are in excellent agreement with the theoretical calculations (right panels). Adapted with permission from [29].

The 3-legged device is preferable for its simplicity and relatively easy fabrication process. Nevertheless, when stress fields with higher magnitude are required it is more convenient to use piezoelectric actuators featuring 6-legs, as demonstrated in [29,128]. In this case, the voltages are applied on pairs of aligned legs so as to avoid lateral displacements at the center of the membrane where the PL characterization is carried out. In this 6-legged device the stress field anisotropy can be tuned up to about 200 MPa (for electric fields on the actuator ranging from -2 to 33 kV/cm) in all the directions (major stress angle) with hydrostatic stress values as large as 350 MPa. This implies that the six-legged devices feature the same capability to control the in-plane stress tensor in nanomembranes as the three-leg device, but the stress magnitude is enhanced by about a factor 2 [29].

**5.2.- Energy-tunable sources of polarization-entangled photons.**

The success of quantum communication technologies for secure long-distance data transmission relies on the successful implementation of quantum repeaters. One of the key ingredients for quantum repeaters is the so-called entanglement swapping among photons generated from distant emitters. As mentioned above, semiconductor QDs are excellent sources of polarization-entangled photons and, since they can be embedded in well-known all-electrically-driven devices, it is conceivable that they could be used to build up a scalable quantum network. However, this idea sets very stringent prerequisites on the emitters, which have to fulfill a long list of requirements. In particular: i) the QDs FSS must be below 1 µeV (a value obtained considering typical lifetime of X transitions) and ii) the emission energy from distant QDs must coincide. As discussed in section 3.1, these requirements imply the deliberate control of the inherent anisotropies of the confining potential in self-assembled semiconductor QDs. Combining two perturbations (*tuning knobs*) to modify the structural symmetries of single QDs, i.e. tuning their FSS value to zero, has been demonstrated in section 4.4 by making use of a vertical electric field in p-i-n diode structures operated in the inverse regime and in-plane strain fields. However, such an approach is not compatible with QLED devices, as the vertical electric field is taken up for tuning the FSS and cannot be easily used at the same time for electrical injection of carriers into the QD. Moreover, there exist specific values of strain and electric field that allows fulfilling (*i*), and any attempt to control the energy of the generated photons (to fulfill ii) restores the FSS and reduces the degree of entanglement in a time-integrated experiment. At this point, it is clear that three independent tuning knobs are needed to build up an energy-tunable polarization-entangled photons: two to tune the FSS to zero and one to tune the emission energy of the photons, as recently demonstrated [19]. These three tuning knobs are given naturally by the tensorial character of the in-plane stress field which provides three independent degrees of freedom ($S_1, S_2, \theta_{s1}$). The corresponding strain Hamiltonian is given by [19]

$$\delta H_s = \bar{\alpha}\bar{S}\tau_0 + \alpha \Delta S \cos(2\theta_{s1})\tau_z + \gamma \Delta S \sin(2\theta_{s1})\tau_x \qquad (8)$$

where the parameters $\bar{\alpha}$, $\alpha$ and $\gamma$ are related to the elastic constants renormalized by the deformation potentials. The hydrostatic part anisotropy and of the stress are given by $\bar{S} = S_1 + S_2$ and $\Delta S = S_1 - S_2$, respectively. Therefore, generally speaking, the three parameters $\bar{S}$, $\Delta S$ and $\theta_{s1}$ are relevant to induce changes in the Hamiltonian. Further details about the underlying theory can be found in Ref. [19].

Note that a given stress field ($S_1$, $S_2$, $\theta_{s1}$) leads to a strain field (deformation) in the material ($\varepsilon_1$, $\varepsilon_2$, $\phi_\varepsilon$) that is given by the product of the elastic compliance tensor and the stress field components, being $\varepsilon_1$, $\varepsilon_2$ the principal major and minor strain, and $\phi_\varepsilon$ the angle between the major strain and the GaAs [110] crystalline direction. As for the case of strain and electric field (see section 4.4) two parameters $\Delta S$ and $\theta_{s1}$ are sufficient to decrease the FSS to zero. This is achieved by aligning the major strain axis ($\phi_\varepsilon$) close to the polarization direction of the exciton emission ($\phi_\varepsilon \sim \phi$) and sweeping $\Delta S$ to reach FSS~0. The theoretical calculations for the case of an InAs QD featuring an exciton polarization angle of $\phi$=16.2° is reported in Figure 26a. In this case, a FSS=0 can be achieved for a major strain angle $\phi_\varepsilon$ =18.5°. The third parameter (hydrostatic part of the stress, $\bar{S}$) is employed to change the emission energy of the exciton without affecting the FSS value, i.e. entanglement. Note that the application of a hydrostatic stress does not change the symmetry of the system (anisotropy is kept constant) but only increases the magnitude of the stress field, i.e. emission energy shifts.

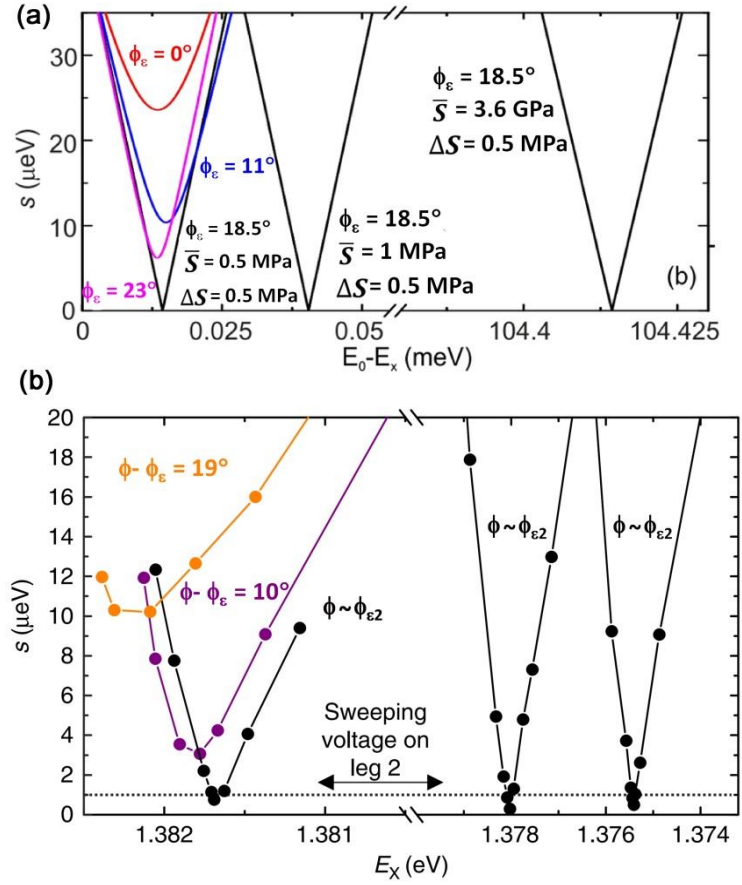

**Figure 26.-** (a) FSS values (s) against the X energy shift ($E_0-E_x$) for an InAs QD with exciton polarization angle ϕ=16.2°. A FSS=0 value is calculated for a major strain angle $\phi_\varepsilon$ =18.5° ($\bar{S}$=0.5 MPa and $\Delta S$=0.5 MPa). The exciton X energy shift is changed by simply varying the hydrostatic stress ($\bar{S}$). Reprinted (figure 2) from [19]. Copyright (2015) by the American Physical Society. (b) Experimental FSS values (s) against the exciton X energy emission for an InAs QD with relative exciton polarization angle with respect to actuating leg-2 direction (major strain angle $\phi_{\varepsilon 2}$) ϕ-$\phi_{\varepsilon 2}$=19°. The exciton X polarization angle can be tuned along the actuating direction of leg-2 by properly tuning the voltage on leg-1 ($V_1$), so that FSS~0 can be reached by sweeping the voltage on leg-2 ($V_2$). The X emission energy can be shifted with different combination of voltages ($V_1,V_2,V_3$) in order to vary $\bar{S}$ while keeping $\Delta S$ and the angle $\phi_{\varepsilon 2}$ constant. Reprinted from [128].

As discussed in the previous section, the in-plane stress tensor can be fully controlled by employing 3-legged or 6-legged micro-machined piezoelectric actuators. In Ref. [128], a 6-legged device is used for the first time to generate energy-tunable polarization-entangled photons from self-assembled InAs QDs embedded in a GaAs nanomembrane. A sketch of the device and the optical picture of the bonded membrane are shown in Figure 27a. The underlying legs and central gap where the measured QDs are located can be clearly distinguished. In order to achieve full control of the FSS in any arbitrary QD, it is necessary to master two different parameters: the polarization angle of the exciton emission (ϕ) and the magnitude of the FSS (s). The former parameter provides information about the orientation of the in-plane exciton dipoles and the associated QD anisotropy [82]. This anisotropy can be visualized with an ellipse whose axes are the two in-plane spin-spin coupling constants [129]. The s and ϕ values are experimentally obtained from polarization-resolved PL spectra on the exciton transition of the QD.

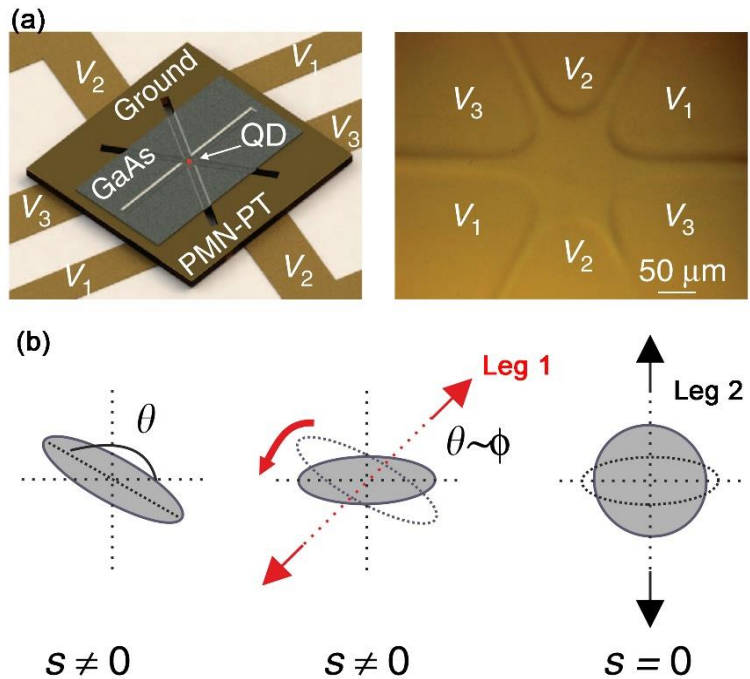

**Figure 27.-** (a) Sketch of a 6-legged device with a GaAs nanomembrane bonded on it (left panel). The PL is measured from QDs located at the central gap between the six legs. Voltages ($V_1,V_2,V_3$) are applied to pairs of aligned legs to avoid lateral displacements of the active structure. An optical picture of the central gap of a 6-legged device is shown (right panel). (b) Schematic sketch of the QD anisotropy for a QD with $s\neq0$ (left panel). The deviations from a symmetric circle indicate the existence of an in-plane anisotropy in the QD confining potential. The anisotropy can be aligned along the actuating direction of leg-2 with leg-1 ($V_1$). Finally, the anisotropies can be fully compensated with leg-2 ($V_2$). Reprinted from [128].

To achieve entangled photons generation with tunable emission energy, a simple three-step procedure can be followed. First, a given voltage is applied on one of the legs to align the polarization angle of the exciton emission along the actuating direction of a second leg. Then, the latter is used to suppress the FSS. Finally, the third leg is used to change the initial QD anisotropy, so that repeating the first two steps results in s=0 at a different energy of the exciton transition. In the case presented in Figure 26b, the polarization angle of the exciton is about ϕ~19° with respect to the [110] axis of the GaAs crystal, which corresponds to the actuating direction of leg-2. In full agreement with the theory, sweeping the voltage across this leg in this condition clearly leads to an anticrossing between the two bright excitonic states (with a minimum value of s~10 μeV). Following the first step of the procedure outlined above,

leg-1 can be used to align the polarization direction of the exciton along leg-2. In this condition, the exciton FSS can be tuned well below 1 µeV, as shown by the black line at the left-hand side of Figure 26b. Control of the exciton energy at zero FSS can be done by applying first a voltage on leg-3 (to modify the strain configuration of the QD and thus the $s$ and $\phi$ values and then using leg-1 to re-align $\phi$ along the actuating direction of leg-2, which is afterwards employed to suppress $s$ at a different energy of the exciton. This procedure can be repeated for different combination of voltages on leg-1, leg-2 and leg-3 and it is demonstrated that the exciton emission can be continuously tuned over a range of ~7 meV at s<1 µeV (Figure 26b). We refer the reader to Ref. [130] for further information. We emphasize that the total tuning range depends on the specific device and on the maximum voltage that can be applied experimentally before the development of dislocations in the PMN-PT actuator or cracks in the bonding layer. Most importantly, the possibility of tuning simultaneously the FSS and entangled photons energy emission by elastic strain engineering through the piezoelectric actuator would allow utilizing the vertical electric field for carrier injection in energy-tunable entangled-light-emitting-diodes (ELEDs).

A similar approach by using a 4-legged device featuring two "tuning knobs" has been reported recently [130]. It should be noted that contrary to the 6-legged device discussed here the possibility of obtaining entangled photons with tunable emission energy for such a 4-legged device is limited to QDs presenting exciton emission polarization orthogonal to one pair of legs. The same consideration applies to the simultaneous use of strain and electric fields to achieve an energy-tunable source of entangled photons [45].

**6.- Conclusions.**

In summary, in this review, we have shown the potential of PMN-PT piezoelectric actuators for tailoring the optical response of a variety of nanomaterials such as semiconductor self-assembled QDs, 2D materials, rare-earth and metal-ions doped thin films and ZnS films presenting piezoelectric properties. The strategy to integrate the optically active materials onto the actuators is of crucial relevance to ensure an efficient strain transfer from the actuator to the attached films/nanomembranes containing the optically active materials. Among all reported data, epoxy-based SU-8 bonding technique offers an impressive 69% strain transfer rate, significantly larger than those reported for epitaxial films on PMN-

PT substrates (up to 40%). Piezoelectric monolithic actuators provide a simple and effective strategy for tuning of the optical emission energy in nanomaterials up to 20 meV in the case of semiconductor QDs, which find interesting applications such as tuning the optical modes into resonance with QDs emission energy in optical cavities for enhanced light emission, tuning the QDs emission energy in nanowire waveguides featuring high light extraction efficiencies or interfacing atomic vapors with QDs for memory applications. Although most of the reported works to date are focused on tuning the optical properties in QDs, the use of piezoelectric actuators has been recently successfully employed for tailoring the optical properties in 2D semiconductor materials and rare-earth/metal-ions doped thin films for applications that might find interesting applications in optoelectronic and quantum devices, as well as fundamental studies. Interestingly, the combination of strain fields with electric fields allows the tuning of the optical properties in QDs – emission energy and FSS – embedded in p-i-n diode structures in an all-electrically driven platform. Additionally, these devices permit utilizing the electric field to further tune the emission energy when the diode is operated in reverse bias, as well as to control independently X and XX emission in QDs and engineering of the excitonic wave function. One of the main drawbacks of this type of devices is the lack of control over the anisotropy of the exerted stress fields, which are fixed after the fabrication of the device. This is especially relevant for the generation of polarization-entangled photons from QDs where a value FSS~0 is mandatory. It has been shown that, upon the application of a strain field with monolithic actuators, this condition can only be satisfied on selected "hero" QDs presenting specific structural anisotropies in the confining potential for the excitons.

Recently, a new class of 2-legged, 3-legged and 6-legged micro-machined piezoelectric actuators have been demonstrated for a full control of the in-plane stress tensor in semiconductor nanomembranes. In particular, the tuning of the emission energy of polarization-entangled photons generated in QDs has been successfully demonstrated. The successful development of such devices mark a technological milestone for the realization of advanced quantum optics experiments not possible to date. A stringent requirement for practical applications is a high light extraction efficiency from the semiconductor nanomembranes containing the QDs. Relative high values up to 60% could be achieved by fabricating monolithically integrated micro-lenses on the nanomembranes, with a gold coating at the bottom acting as an optical mirror, by in-situ lithography followed by etching [131]. In addition, the possibility of introducing combined strain and electric fields (reversed bias) in stretchable diode-like nanomembranes integrated on 6-legged devices would allow all-electrically-driven sources of energy-tunable single photons entangled in polarization, with broadband light extraction enhancement. We envision a handful of possibilities such as teleportation of photon quantum states through entanglement swapping in pair of photons emitted from distant QDs sources of entangled single photons for applications in long distance quantum communication schemes. The possibility of interfacing photons with warm atomic vapors would allow for the storing of photons for quantum memories in quantum repeaters networks [132].

More generally, the piezoelectric devices presented in this review represent a versatile approach to perform fundamental studies, in vacuum conditions and at cryogenic temperatures, to provide fundamental knowledge and understanding of the strain effects on new technologically interesting materials. A prominent example is the broad family of semiconductor 2D materials including superconductors, semiconductors or topological insulators. Moreover, micro-machined actuators could be used to study the effect of isotropic/anisotropic biaxial stress fields in an unprecedented way on the anisotropic electrical conductivity, tunability of hyperbolic plasmons and optical response of a variety of 2D materials featuring in-plane structural anisotropies [133–135]. Finally, it should be noted that due to the versatility of the fabrication processes presented in this review following lift-off techniques and/or epitaxial growth, it should be possible to integrate a vast majority of materials onto piezoelectric devices, which opens new avenues for the exploitation of ESE of materials.

**Acknowledgements.-**
This work was financially supported by European Union's Horizon 2020 research and innovation programme (SPQRel grant agreement no. 679183), European Union Seventh Framework Programme (FP7/2007–2013) under grant agreement no. 601126 (HANAS) and Austrian Science Fund (FWF): P 29603. We thank Georgios Katsaros, Daniel Huber and Marcus Reindl for fruitful discussions. Alma Halilovic, Stephan Bräuer and Ursula Kainz are acknowledged for technical support.